\documentclass[12pt,onecolumn]{article}

\usepackage{mathtools}
\usepackage{cuted}

\usepackage{booktabs,siunitx}
\usepackage{amsbsy}
\usepackage{amsfonts}
\usepackage{amssymb}
\usepackage{cite}
\usepackage{braket}
\usepackage{bm}
\usepackage{leftidx}
\usepackage{enumerate}
\usepackage{graphicx}
\usepackage{epsfig}
\usepackage{color}
\usepackage{nicefrac}
\usepackage{mathrsfs}
\usepackage{bbm}
\usepackage{amsmath}
\usepackage[colorlinks]{hyperref}
\usepackage{hyperref}
\usepackage{mathrsfs}
\definecolor{gruen}{rgb}{0,0.625,0}

\newlength{\TZ}
\newcommand{\To}{\rightarrow}
\newcommand{\s}{\mathscr{S}}

\newcommand{\BEQ}{\begin{equation}}     % Gleichungen Anfang ..
\newcommand{\BEA}{\begin{eqnarray}}
\newcommand{\BD}{\begin{displaymath}}
\newcommand{\EEQ}{\end{equation}}       % .. und Ende
\newcommand{\EEA}{\end{eqnarray}}
\newcommand{\ED}{\end{displaymath}}
\newcommand{\bb}{\begin{eqnarray}}
\newcommand{\ee}{\end{eqnarray}}

\newcommand{\D}{{\rm d}}                % gerades d fuer Ableitungen
\renewcommand{\vec}[1]{\boldsymbol{#1}} % Vektoren fettgedruckt

\newcommand{\blue}{\textcolor{black}}
\catcode`\@=11
\def\numberbysection{\@addtoreset{equation}{section}
        \def\theequation{\thesection.\arabic{equation}}}
\numberbysection

\begin{document}

\begin{titlepage}

% \vspace*{.5cm}
\begin{center}
{\Large \bf Dynamical off-equilibrium scaling across magnetic \\[.25cm]
first-order phase transitions}
\end{center}

\vskip 2.0 cm
\centerline{{\bf Stefano Scopa}$^{a,}$\footnote{email: stefano.scopa@univ-lorraine.fr,
ORCid:  0000-0001-7638-8804} and {\bf Sascha Wald}$^{b,}$\footnote{email: swald@sissa.it,
ORCid:  0000-0003-1013-2130} }
\vskip 0.5 cm
\begin{center}
$^a$
Laboratoire de Physique et Chimie Th\'eoriques,  UMR  CNRS  7019,
Universit\'e de Lorraine BP 70239, F-54506 Vandoeuvre-l\`es-Nancy Cedex,  France
\\ \vspace{0.5cm}
$^b$ SISSA - International School for Advanced Studies and INFN,\\
via Bonomea 265, I--34136 Trieste, Italy
\vskip 0.5 cm
\today
\end{center}

\vskip 1.0 cm
\begin{abstract}
We investigate the off-equilibrium dynamics of a classical spin system with 
$O(n)$ symmetry in $2< D <4$
spatial dimensions and in the limit $n\to \infty$. The system is set up in an 
ordered equilibrium state 
and is subsequently driven out of equilibrium by slowly varying the external 
magnetic field $h$ across 
the transition line $h_c=0$ at fixed temperature $T\leq T_c$. We 
distinguish the cases $T = T_c$ where the magnetic transition is continuous and 
$T<T_c$
where the transition is discontinuous. In the former case, we apply a standard 
Kibble-Zurek 
approach to describe the non-equilibrium scaling
and formally compute the correlation functions and scaling relations. For the 
discontinuous transition we develop a scaling theory which builds on the 
coherence length rather than the correlation length since 
the latter remains finite for all times. Finally,  we derive the off-equilibrium 
scaling relations for the hysteresis loop area during a round-trip protocol
that takes the system across its phase transition and back. Remarkably, our 
results are valid beyond the large-$n$ limit.
\end{abstract}

\vfill
%\noindent

%PACS numbers: 05.20.-y, 05.10.Gg, 05.70.Ln, 64.60.De \\~\\
%\textcolor{black}{\tt pr\'eliminaire \hfill \today}

\end{titlepage}
 \onecolumn\tableofcontents
%
%  \twocolumn
%

\setcounter{footnote}{0}
\section{Introduction}
%\textcolor{black}{RESTORED}
The study of equilibrium statistical mechanics and especially of critical phenomena has lead to a refined physical 
understanding of complex, interacting systems and their collective behaviour \cite{Amit84,Card96,Sach99,Nish11,Wipf13}. In the 
last decades, the study of 
equilibration processes and out-of-equilibrium dynamics has gained significant attention in order
to complement the equilibrium studies, to understand how systems behave far from equilibrium and how thermalisation may 
occur \cite{Henk09,Henk10,Cugl03,Tauber}. 

\begin{figure}[ht]
 \centering
  \includegraphics[width=.6\textwidth]{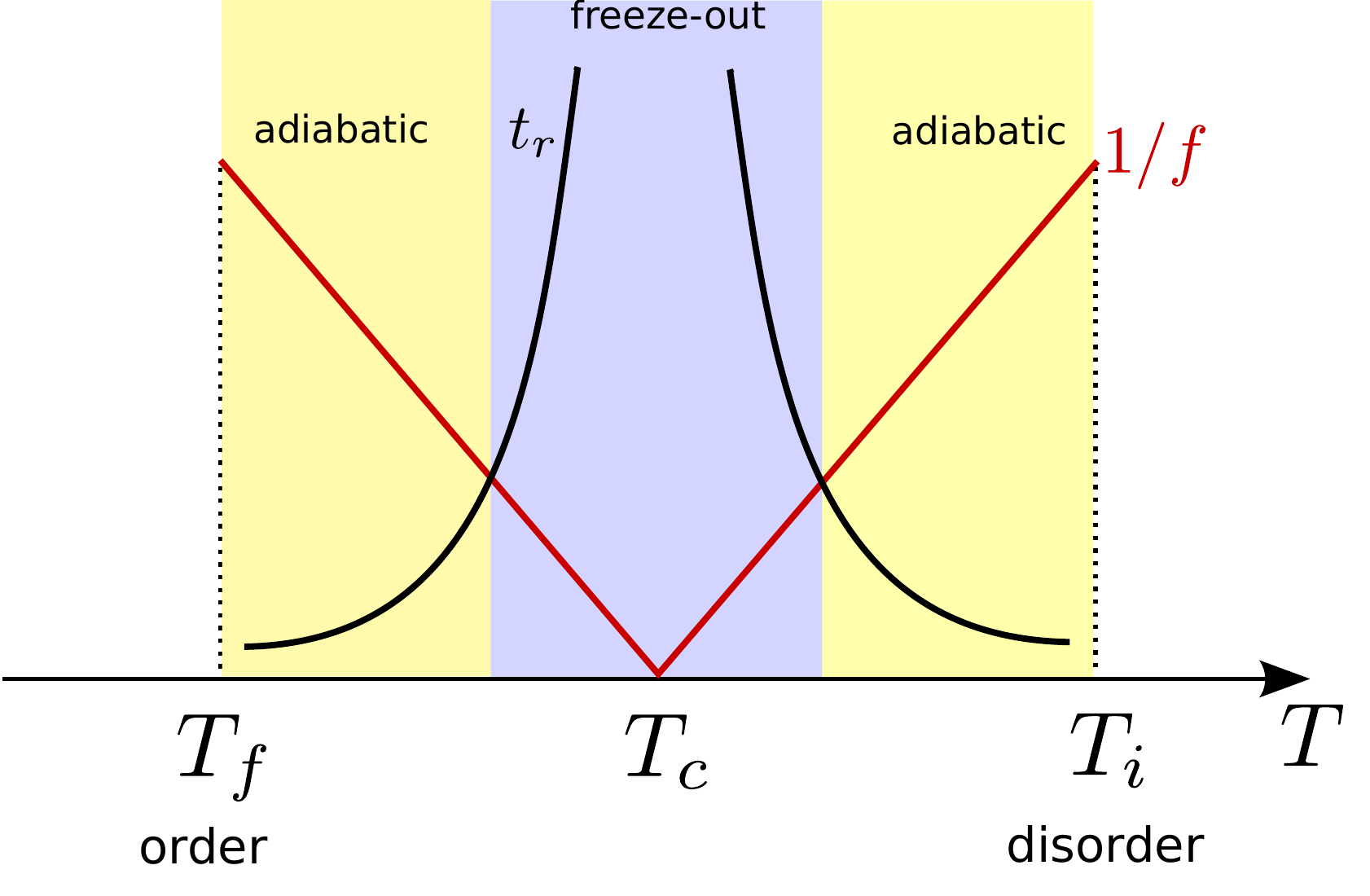}
\caption{\small
Qualitative illustration of the {\sc kz} mechanism. We prepare the system initially in equilibrium at a certain 
temperature $T_i>T_c$ in the disordered phase. We then vary the temperature in time at a finite rate $f$ across the 
phase boundary up to a final value $T_f<T_c$. As long as the relaxation time
$t_r < 1/f$ the system adapts to the temperature variation and adiabatically follows the quench protocol (yellow regions).
In the vicinity of the transition point, the divergence of $t_r$ leads to a time $\tau$ where
$t_r = 1/f$ after which the system cannot adjust to the temperature variation anymore and falls collectively out of equilibrium 
(blue region). The system remains frozen in this region until $t_r$ decays again in the ordered phase.
\textcolor{black}{This phenomenological picture can be better understood by means of \textit{finite time scaling}
\cite{Zhong2014,Zhong2010,Zhong2016}.
}}  
\label{fig:KZM}
\end{figure}

A standard way to produce non-equilibrium situations is by studying {\it quench protocols} where either
an external parameter (e.g. the temperature) or a Hamiltonian parameter (e.g. the interaction 
strength) is varied in time across a \textit{phase transition}. By means of such 
protocols, one is able to drive a system through different regions of the phase diagram and to investigate
relaxation and thermalisation properties \cite{Cugl03,Henk09,Henk10,Stru78,Bray94,Cates00,Godr02,Paes03,Breu07,Scha14,Schmalian15,Mar15,Chio17,Schmalian14,Wald16,Wald18a,Wald18b,Gar04}.
Most of the dedicated literature refers to quench protocols across continuous phase transitions, see e.g.
\cite{Godr02,Paes03,Wald16,Wald18a,Godr00b,Godr13} but this list is far from being exhaustive. 
If the 
driving across such a transition is performed slowly (in a sense that will be specified later on),
these quench protocols are described by the
{\it Kibble-Zurek} ({\sc kz}) {\it scaling theory} \cite{Kibble,Kibble2,Kibble3,Zurek-85,Zurek-96}, which explains the 
formation of topological defects occurring after the quench. The main idea of this approach 
is illustrated in figure~\ref{fig:KZM}. The {\sc kz} scaling theory has been tested in a variety of experiments \cite{Donatello-16,Lamporesi-13,Pyka-13,Corman-14,Cui-16} where it has been shown to describe the off-equilibrium dynamics near transitions well and 
especially to predict experimental data for the density of topological defects accurately. In fact, the {\sc kz} theory has 
proven itself to be of great use for the description of non-equilibrium properties especially since in general, more complicated
methods are used to analyse such scenarios \cite{Cala02,Godr00b,Wald18c,Cala05}.
Therefore the {\sc kz} theory has been {\it inter alia} extended to quantum phase transitions \cite{Zurek-05,QKZ, QKZ2,qkz3,rev-off-eq1,rev-off-eq2},
where experiments with ultracold atoms in optical lattices provide an ideal platform for applications of the theory \cite{KZ-cold_atoms,KZ-cold_atoms2,DelCampo-11,Scopa,
Land16,Hruby18,Landini18}.

\begin{figure}[ht]
 \centering
\includegraphics[width=.6\textwidth]{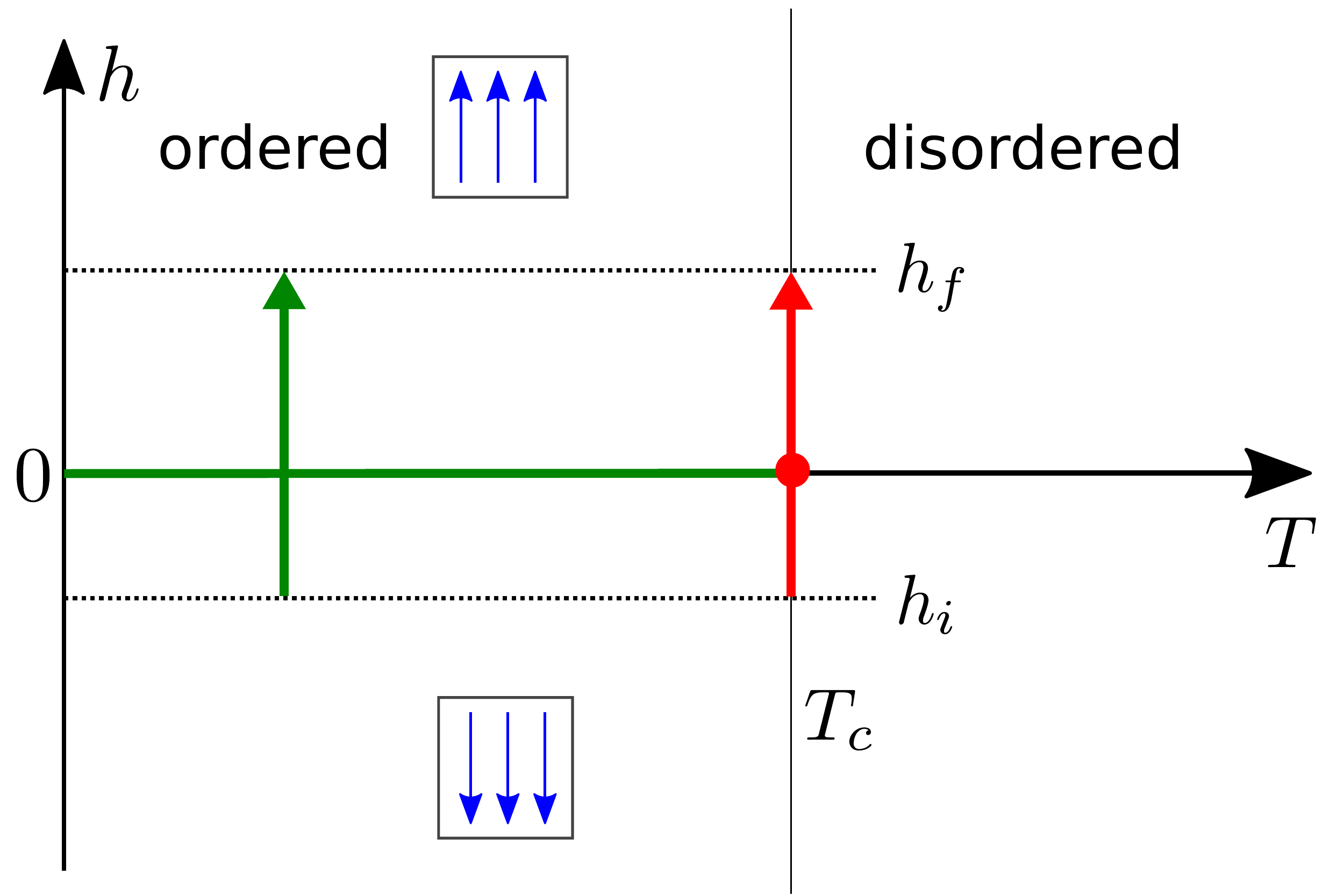}
\caption{\small
Schematic representation of the phase diagram for a generic spin system. The red dot indicates a
continuous phase transition while the green line indicates a discontinuous transition. The corresponing arrows
show qualitatively the quench protocols we shall study.}
\label{fig:phase}
\end{figure}

%At first this approximation may seem very pragmatic or even drastic but indeed the {\sc kz} approach has been tested in a variety 
%of experiments \cite{Donatello-16,Lamporesi-13,Pyka-13,Corman-14,Cui-16} where it has been shown to describe the off-equilibrium dynamics near transitions well.
%Especially quantitative predictions for the density of topological defects after the quench reflect experimental data well.
%In fact, the {\sc kz} theory has proved itself to be of great use and therefore has been inter alia extended to quantum phase transitions \cite{Zurek-05,QKZ, QKZ2,qkz3,rev-off-eq1,rev-off-eq2},
%where experiments with ultracold atoms in optical lattices provide an ideal platform for applications of the theory \cite{KZ-cold_atoms,KZ-cold_atoms2,DelCampo-11,Scopa,
%Land16,Hruby18,Landini18}.
The aim of this work is instead to extend the {\sc kz} scaling theory to quench protocols across a {\it first-order transition}
({\sc fot}). To do so, we consider a generic spin system which is known to show a phase transition from a magnetic {\it down} to 
{\it up} order, driven by an external magnetic field $h$ at
a fixed temperature $T\leq T_c$, see figure~\ref{fig:phase}. The nature of this transition depends on the temperature
at which the magnetic transition is driven.
\begin{itemize}
 \item At $T=T_c$ the transition is {\it continuous} and the standard {\sc kz} scaling theory applies.
 %will describe this transition as illustrated in figure~\ref{fig:KZM}.
 \item For $T<T_c$ the transition is {\it discontinuous} and therefore the  system correlation length remains finite 
 at all times.
\end{itemize}

In the latter case,  a {\it new} description is needed since the {\sc kz} theory is built on the fact that the correlation length $\xi$ diverges at the critical point. Nevertheless, an equilibrium scaling theory has been developed for {\sc fot}s \cite{Nienhuis-75,Fisher-82,Privman-83,Binder}
by replacing the correlation length with the {\it coherence length} $\xi_h$, which corresponds to the typical domain size of ordered clusters \blue{in minimal energy configuration. As soon as the discontinuous transition is approached, 
the two ordered phases become energetically indistinguishable, leading to a divergence of $\xi_h$. 
This divergence is physically reflected in magnetic systems by the long-range order arising in the spin-spin correlation function which asymptotically ($|\vec{x}-\vec{x}^{\prime}|\To\infty$) approaches the value of the squared magnetisation of the system \cite{Fisher-82}. Notice that the correlation length $\xi$ is instead defined by the connected part of the spin-spin correlator $G(\vec{x}, \vec{x}^{\prime}) \propto \exp\left(-|\vec{x}-\vec{x}^{\prime}|/\xi\right)$ and remains finite in the non-critical regime.}

%\textcolor{black}{which is \blue{defined through} the \blue{two-point} correlation function
%\begin{equation}
% G(\vec{x}, \vec{x}^{\prime}) \propto \frac{ e^{-|\vec{x}-\vec{x}^{\prime}|/\xi}}{|\vec{x}-\vec{x}^{\prime}|^{D-2+\eta}} \ ,
%\end{equation}
%}
% \textcolor{black}{This is not the case for a {\sc fot} since such a transition 
%point seperates two \textit{non-critical} phases.}
%Nevertheless, an equilibrium scaling theory has been developed for {\sc fot}s \cite{Nienhuis-75,Fisher-82,Privman-83,Binder}
%by replacing the correlation length with the {\it coherence length} $\xi_h$, which corresponds to the typical domain size of ordered clusters. \textcolor{black}{The quantity $\xi_h$ is related to the energy landscape near
%the transition and its divergence signals the occurance of long-range order.}

%
In this paper, we shall naturally extend this {\sc fot} scaling theory to the non-equilibrium 
case {\it \`a la} {\sc kz} and apply it to magnetic quench protocols in the ordered region as shown in figure~\ref{fig:phase} (green).
\textcolor{black}{This treatment is complementary to 
previous renormalisation group studies \cite{Zhong1995,Zhong2005,Zhong2006} and recent numerical evidence of dynamical scaling across {\sc fot}s \cite{Zhong2018}.}

% 
% 
% being 
% continuous at $T=T_c$ and discontinuous for $T<T_c$, as shown in . By varying the external magnetic
% field,  we are able to explore the phase diagram and to investigate the non-equilibrium dynamics arising close to the point 
% $h=0$.

% While at $T=T_c$ a standard {\sc kz} scaling applies, for values of the temperature $T <T_c$ 

A theoretical framework for the description of generic spin systems is provided by the $O(n)$ model which is routinely cast as a field theory with the 
$n$-component vector field $\vec{\phi}(\vec{x})$ of unit 
norm and the action (see e.g. \cite{ZinnJustin})
% 1-Column form
 \begin{equation}\label{eq:S4}
 S[\vec{\phi}] = \frac{n}{2}\; \int \D \vec{x}\left[ (\nabla \vec{\phi}(\vec{x}))^2 + r\vec{\phi}^2(\vec{x})
 +u (\vec{\phi}^2(\vec{x}))^2 -2\vec{h}\, \vec{\phi}(\vec{x})\right] 
\end{equation}
%2-column form
% \begin{align}\nonumber
%  S[\vec{\phi}] = \frac{n}{2}\; \int \D \vec{x}\big[ (\nabla \vec{\phi}(\vec{x}))^2 + r\vec{\phi}^2(\vec{x})\\
%  +\frac{u}{12} (\vec{\phi}^2(\vec{x}))^2 -2\vec{h}\cdot \vec{\phi}(\vec{x})\big]
% \end{align}
in $2<D<4$ spatial dimensions. Here, $u>0$ is a positive constant, $\vec{h}$ describes the \textit{external magnetic field} and 
$r=r(T)$ is the thermal coupling constant. This model includes the celebrated Ising model ($n=1$), the {\sc xy} model ($n=2$) and the Heisenberg model ($n=3$) as 
special cases \cite{Stanley-O(N)}. In the particular case where $n\to \infty$, the bulk critical behaviour
reduces to the one of the spherical model \cite{Stanley} and allows analytic investigation \cite{Baxt82,Oliv06,Vojta96,Berl52,Lewi52,Wald15,Henk84a}.
% We shall refer to this scenario as the ``large-$n$ limit''.
%Statistical mechanics in and out of equilibrium has always
%profited from such analytic solutions in order to better understand underlying physical mechanism or to benchmark existing numerical
%approaches \cite{Baxt82}. This is our motivation to be mainly concerned with this limit to which we refer as ``large-$n$ limit''.
The analyticity in the large-$n$ limit arises due to the central limit theorem which implies that
self-averaged $O(n)$ symmetric quantities are normally distributed for $n\to\infty$. 
This allows us to replace \cite{LargeN-rev}
\begin{equation}
 (\vec{\phi}^2(\vec{x}))^2 \to \left<\vec{\phi}^2(\vec{x})\right>\vec{\phi}^2(\vec{x})
\end{equation}
and cast the action in eq~(\ref{eq:S4}) quadratically
% 1-column
% \begin{equation}
%  S[\vec{\phi}] \to \frac{n}{2}\; \int \D \vec{x}\left[ (\nabla \vec{\phi}(\vec{x}))^2 + m^2\vec{\phi}^2(\vec{x})-2\vec{h} . \vec{\phi}(\vec{x})\right], 
% \label{eq:Slargen}
% \end{equation}
\begin{align}
 S[\vec{\phi}]= \frac{n}{2}\! \int\! \D \vec{x} \big[ (\nabla \vec{\phi}(\vec{x}))^2\! +\! m^2\vec{\phi}^2(\vec{x})
\! -\!2\vec{h} \, \vec{\phi}(\vec{x})\big]
\label{eq:Slargen}
\end{align}
with the {\it effective mass}
\begin{equation}
 m^2 = r +u\left<\vec{\phi}^2(\vec{x})\right>.
\label{eq:meff}
\end{equation}

For this specific model, the $O(n)$ model at large $n$, we shall % test the off-equilibrium scaling Ansatz and 
derive the structure of the off-equilibrium 
scaling functions of the magnetisation and of the transverse correlation functions. The latter are self-consistently related 
through an equation of state which describes - roughly speaking -  the time-evolution of the magnetisation as a 
rotation generated by the dissipation of the initial equilibrium magnetisation into the transverse field modes 
(see section~\ref{ssec:t<tc} for further details). We shall also provide a scaling prediction for the 
dissipated magnetic work $W$ during a round-trip protocol  which takes the system across the {\sc fot} and back.
If we call the quench time scale $t_s$, one obtains in $D=3$ spatial dimensions
\begin{align}
% \label{heis-work-tc}
W \propto \begin{cases}
           t_s^{-2/3}, \hspace{.25cm} \text{for} \hspace{.25cm} T=T_c\\[.5cm]
           t_s^{-1/2}, \hspace{.35cm}\text{for} \hspace{.25cm}    T<T_c
          \end{cases}
\ .
\end{align}
Remarkably, these results apply beyond the large-$n$ limit with $2/3\approx 0.66$, as 
can be seen by a comparison with numerical results by {\it Pelissetto and Vicari} \cite{Vicari-off-16}.
To complete our analysis, we shall numerically investigate the dynamics in the large-$n$ limit and explicitly test our 
scaling predictions.

The paper is organised as follows: After having set up the model in this introduction we turn to 
its general dynamical description in section \ref{sec:dyn}. Here, we specify the magnetic quench protocol that we shall study and we determine the set
of dynamical equations that describes the $O(n)$ model self-consistently for $n\to\infty$. In section \ref{sec:scaling} we 
then derive the dynamical
scaling theory for the magnetic quench. 
We explicitly distinguish between (i) $T=T_c$ where we verify the {\sc kz} scaling and (ii) $T<T_c$ where we develop a new
out-of-equilibrium scaling theory.
%Here we shall distinguish between a quench across the continuous transition for 
%which a {\sc kz} scaling applies and one across the {\sc fot}  for which we develop a scaling theory inspired by the {\sc kz} 
%approach. 
Finally, we apply these results in section \ref{sec:hys} to a round-trip protocol for which we then derive the 
scaling behaviour of the hysteresis area and of the magnetic work performed over the cycle.
%that indicates the non-equilibrium properties of our model. 
We then briefly summarise our results and conclude the paper. Several technical aspects are described in the appendix.

\section{Dynamical description of the $O(n)$ model}\label{sec:dyn}
\begin{figure}[ht]
 \centering
 \includegraphics[width=.4\textwidth]{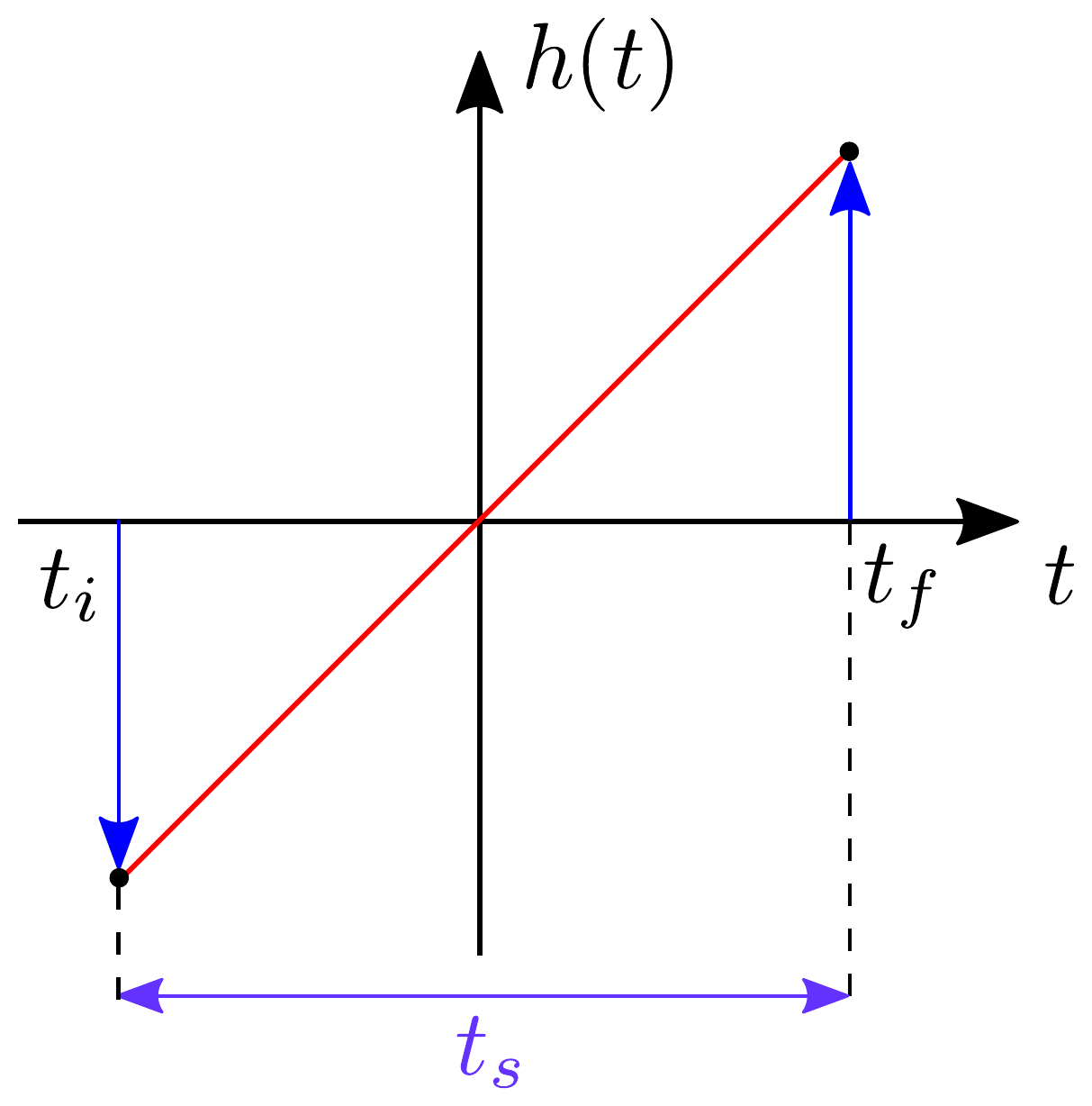}
 \caption{\small Schematic representation of the magnetic quench protocol~(\ref{eq:quench}). At initial time $t_i<0$ the system is in thermal
 equilibrium with the external magnetic field $h(t_i)<0$. This field is then linearly driven through the magnetic
 transition point $h_c = 0$ on a time scale $t_s$ until it reaches its final value $h(t_f) > 0$.}
 \label{fig:ramp}
\end{figure}

We want to describe the dynamics of the system~(\ref{eq:Slargen}) at and below the critical temperature $T_c$
when the external magnetic field is varied in time across the value $h_c = 0$. We choose the 
{\it linear}\footnote{The extension to non-linear protocols is straightforward, see e.g. \cite{Sondhi-12}.} ramp sketched 
in figure~\ref{fig:ramp} along a fixed direction $\vec{e}_1$
which takes the system from a {\it down} order at initial time $t_i<0$ to an {\it up} order at the final time $t_f>0$, i.e.
\begin{equation}\label{eq:quench}
\vec{h}(t)= \frac{t \ \vec{e}_{1}}{t_f-t_i} = t /t_s\, \vec{e}_{1} \ .
\end{equation}
Here, $t_s=t_f-t_i$ defines the {\it time-scale of the quench}. 
Within this convention, the critical value $h_c = 0$ is reached at time $t=0$ which is but a convenient choice.

The dynamics of the components of the vector field is given by a Langevin equation
% In order to describe such a dynamics, 
% we have also to specify how the system relaxes. We consider the
%a relaxational dynamics given by a standard Langevin equation  for the components of the vector field
% and choose our system of coordinates in a fashion such that the first component aligns with $h$
%
\begin{equation}\label{dynamics}
 \partial_t \phi_{a} (\vec{x},t) = -\frac{\delta}{\delta\phi_{a}} S[\vec{\phi}] + \zeta_a(\vec{x},t) \ ,
\end{equation}
where $\zeta_a(x,t)$ is a Gaussian white noise with zero mean, i.e.\footnote{The damping rate is set
to unity here. The variance is set to $2$ in order for the long-time limit of the two-point function to correctly reproduce 
the equilibrium value when $\vec{h}(t) = cst$.}
%1-COLUMN
% \begin{align}
%  \left<\zeta_{\alpha}(\vec{x},t)\right> = 0\ , \hspace{1cm} 
% \left<\zeta_\alpha(\vec{x},t)\zeta_\beta(\vec{y},t')\right> = 2\delta_{\alpha,\beta} \delta(\vec{x}-\vec{y}) \delta(t-t')
% \end{align}	
\begin{align}\label{noise}
 \big<\zeta_{a}(\vec{x},t)\big> &= 0\ ,\\
\big<\zeta_a(\vec{x},t)\zeta_b(\vec{y},t')\big>  &= 2\delta_{a,b} \delta(\vec{x}-\vec{y}) \delta(t-t') \ .
\end{align}
%over which averages of time-dependent quantities are taken. 
The dynamics of the system is more involved than the one of a standard Gaussian theory due to eq~(\ref{eq:meff}) which 
has to be taken into account self-consistently.
% 
% 
% In order to describe the dynamics of the system that takes eq~(\ref{eq:meff}) self-consistently into account we introduce 
% some notation. 
% 
To do so, we first introduce the time-dependent magnetisation 
\begin{equation}
\big<\phi_{a}(\vec{x},t)\big>= \delta_{1,a} \, M(t)
\end{equation}
as order parameter of the transition. Moreover, we define
the longitudinal and orthogonal (connected) correlation function as
% 1-column
\begin{align}
 G_{||}(x-y,t) &\equiv \left< \Big(\phi_{1}(\vec{x},t)-\braket{\phi_{1}(\vec{x},t)}\Big)\Big(\phi_{1}(\vec{y},t)
 -\braket{\phi_{1}(\vec{y},t)}\Big)\right>\ ,\\[.25cm]
 G_{\perp}(x-y,t)&\equiv \left< \Big(\phi_{a}(\vec{x},t)-\braket{\phi_{a}(\vec{x},t)}\Big)\Big(\phi_{a}(\vec{y},t) -\braket{\phi_{a}(\vec{y},t)}\Big)\right>, \ a>1
\end{align}
% \begin{align}\nonumber
%  G_{||}(x-y,t) &:= \big< \big(\phi_{1}(\vec{x},t)-\braket{\phi_{1}(\vec{x},t)}\big)\\
% &\hspace{-.5cm}\times\big(\phi_{1}(\vec{y},t)-\braket{\phi_{1}(\vec{y},t)}\big)\big>\\[.5cm] \nonumber
%  G_{\perp}(x-y,t)&:= \big< \big(\phi_{a}(\vec{x},t)-\braket{\phi_{a}(\vec{x},t)}\big)
% \\
% &\hspace{-.5cm}\times\big(\phi_{a}(\vec{y},t)
% -\braket{\phi_{a}(\vec{y},t)}\big)\big>, \ a>1
% \end{align}

Due to translational invariance, the dynamics is straightforwardly described in Fourier space and it is easy
to show that it is governed by the following set of equations \cite{Mazenko-85}
\begin{align}\label{dyn-obs}
\frac{\D}{\D t}M(t)&=-m^2(t)  M(t) + h(t)\ , \\[.25cm]
\partial_t G_{\perp}(\vec{q},t)&=-2(m^2(t)+q^2) G_{\perp}(\vec{q},t) +2 \ ,
\label{dyn-obsG}
\end{align}
where the time-dependent effective mass $m(t)$ is defined through the {\it equation of state}
\begin{equation}\label{time-eq-state}
m^2(t)= r+u\bigg[M^2(t) +\int_q G_{\perp}({\bf q},t)\bigg].
\end{equation}
Here, we used the shorthand $\int_q = \int^\Lambda \D {\bf q}/(2\pi)^d $ with the momentum cut-off $\Lambda$.
The dynamical eqs~(\ref{dyn-obs},\ref{dyn-obsG}) can be formally solved as follows
% Assuming that the system is initially in 
% equilibrium, we obtain for the magnetisation
%
% 1 column
% \begin{strip}
% \begin{minipage}{.5\textwidth}
% \ \hrule
% \end{minipage}
 \begin{subequations}
\begin{align}\label{eq:M-t}
M(t)&=M_0\, \exp\bigg[-\int_{t_i}^t \D u\, m^2(u)\bigg]  
+ \int_{t_i}^t \D u \, h(u)\, \exp\bigg[-\int_{u}^t \D s \, m^2(s)\bigg],
\\[.25cm]
\label{eq:GT-t}
G_{\perp}(\vec{q},t)& =2 \int_{t_i}^t \D u \exp\bigg[-2\int_{u}^t \D s \, \big(\vec{q}^2+m^2(s)\big)\bigg]  \ ,
\end{align}
\end{subequations}
%
% \end{strip}
%
%
% \begin{subequations}
% \begin{align}\nonumber
% M(t)&=M_0\, \exp\bigg[-\int_{t_i}^t \D t'\, m^2(t')\bigg]\\\label{eq:M-t}
%  + &\int_{t_i}^t \D t' \, h(t')\, \exp\bigg[-\int_{t'}^t \D t'' \, m^2(t'')\bigg],
% \\[.5cm]
% \label{eq:GT-t}
% G_{\perp}(\vec{q},t)&=2\hspace{-.1cm}\int_{t_0}^t\hspace{-.1cm} \D s\, \exp\bigg[\hspace{-.1cm}-
% \hspace{-.1cm}2\hspace{-.1cm}\int_{s}^t\hspace{-.1cm} \D u \, \big(\vec{q}^2\hspace{-.1cm}+m^2(u)\bigg] 
% \end{align}
% \end{subequations}
%
with the initial equilibrium magnetisation $M_0$. In the following, we shall use these formal solutions
(\ref{eq:M-t},\ref{eq:GT-t}) together with the equation of state (\ref{time-eq-state}) in order to describe the model 
out of equilibrium.
%\textcolor{black}{Notice that the time-evolution of the correlations above is determined not merely by the instantaneous value of the 
%external field but depends also on the values of the external field and of the correlators (through $\eqref{time-eq-state}$) 
%during the previous instants of time. This kind of memory effects are a clear signature of a non-equilibrium dynamics.}
%

%###############################################################
%#######Dynamical scaling theory################################
%###############################################################
\section{Dynamical scaling theory across phase transitions}
\label{sec:scaling}
In this section we develop a scaling theory {\it \`a la} {\sc kz} for {\sc fot}s in order to describe the magnetic quench 
specified in eq~(\ref{eq:quench}). 
%
%To do this, we have to distinguish between the cases
%\begin{itemize}
% \item $\beta=\beta_c$ where the system undergoes a continuous magnetic transition,
% \item $\beta>\beta_c$ where the system shows a {\sc fot}.
%\end{itemize}
% at thermal criticality $T=T_c$
% and one below $T_c$ since the former the system undergoes a continuous transition while in the latter it shows a {\sc fot}.
% (where the system is driven across a 
% continuous phase transition) and the case $T<T_c$ which regards a quench across a {\sc fot}.\\
%
We shall though first start with the instructive case $T=T_c$ in section~\ref{ssec:tc} in order to illustrate 
the standard {\sc kz} theory for continuous transitions.
We then turn to the case $T<T_c$ in section~\ref{ssec:t<tc} where we shall develop the non-equilibrium scaling theory for 
{\sc fot}s.

Along with our scaling analysis, we shall provide numerical solutions of the dynamical equations of the $O(\infty)$ model
$(\ref{dyn-obs},\ref{dyn-obsG},\ref{time-eq-state})$ and we shall use these results to check our scaling predictions.
The numerical calculations will be carried out in $D=3$ spatial dimensions and with the normalisation $u=1$ which implies
$r_c \simeq -0.051$ \cite{Mazenko-85,Sondhi-12}. For further details on the numerical method, see appendix \ref{numerics}.

\subsection{Scaling theory for the continuous transition ($T=T_c$)}
\label{ssec:tc}

% We start our analysis by considering the continuous phase transition that occurs as a function of the magnetic field at fixed
% temperature $T=T_c$. 
In this case the standard {\sc kz} scaling theory \cite{Kibble,Zurek-85} describes the universal scaling behaviour 
of the dynamics driven by the protocol~(\ref{eq:quench}).
First, we have to express the correlation length $\xi$ close to the critical point $h_c=0$ as 
a power-law of the control parameter \cite{Sondhi-12}
% For a system initially prepared at $T=T_c$ and $h(t_i)<0$, the correlation length
% shows a power-law behaviour close to the critical point $h=0$ that can be expressed in terms  of the control parameter
\begin{equation}
\xi(t) \sim |h(t)|^{-\nu_h}, \; \; h\To 0
\label{eq:corrlength}
\end{equation}
% where the symbol $\sim$ denotes a scaling relation and 
with the equilibrium critical exponent\footnote{The {\sc rg} critical exponent 
for the $O(\infty)$ model is $\eta=0$, see e.g. \cite{ZinnJustin-VectorN3,Berl52}.}
\begin{equation}
\nu_h=\frac{1}{d_h}=\frac{2}{D+2} \ .
\end{equation}
\noindent
% The time-scale over which the system adapts
From $\xi$ in eq~(\ref{eq:corrlength}) we can define the typical time scale on which the system adapts 
to the variation of the magnetic field via $t_{\text{ad}}(t)= \xi/\dot{\xi}$ and compare it to the relaxation time 
associated with $\xi$ via $t_{\text{r}}(t) \sim \xi^z$, where $z=2$ is the dynamical critical exponent \cite{Cardy}. These 
time scales do compete during the quench as the system
tries to relax towards equilibrium {\it and} to follow the quench protocol simultaneously. The Ansatz 
that underlies the {\sc kz} approximation is that the system manages to equilibrate and to follow the quench adiabatically
as long as $t_{\text{ad}}< t_{\text{r}}$, compare figure~\ref{fig:KZM}. 
In the opposite situation $t_{\text{ad}}> t_{\text{r}}$ the system cannot adapt to external changes anymore and is % in a zeroth order approximation
assumed to freeze out.
% 
% 
% 
% On the other hand, it is also well-known that
% the time of relaxation $t_{relax}$ of fluctuation modes becomes large close to the criticality. Its growth has power-law 
% behaviour $t_{relax}(t) \sim (\xi(t))^z$ with a dynamical exponent $z=2$. From the interplay of these two competiting effects 
% we can conclude that the system can be considered in equilibrium at each instant of time for which the condition 
% $t_{adapt}(t)<t_{relax}(t)$ is satisfied. In contrast, when $t_{adapt}(t)>t_{relax}(t)$, the system cannot reach to adapt 
% itself to the external variations and goes out of equilibrium.
% 
The crossover time $\tau$ at which the system falls collectively out of equilibrium is then given by
\begin{equation}\label{KZ-t}
t_{\text{ad}}(\tau)\stackrel{!}{=}t_{\text{r}}(\tau) \; \Rightarrow \; \tau=t_s^{\frac{z\nu_h}{1+z\nu_h}} \ .
\end{equation}
From this, we define, \textcolor{black}{in analogy to the equilibrium correlation length}, a \textit{characteristic length scale} $\ell$ via the dynamical exponent $z$, i.e.
% the correlation length of the system at the moment when the freeze-out occurs
\begin{equation}\label{ell}
\ell=\tau^{1/z}=t_s^{\frac{\nu_h}{1+z\nu_h}} \ .
\end{equation}
This length scale encodes the characteristic distance on which the system is correlated and thus allows us to study
the non-equilibrium regime in a scaling limit. In this {\sc kz} scaling regime, the quench is assumed to be slow $t_{s}\to \infty$
while $t/\tau$ and $q\, \ell$ are kept constant.
% Notice that the length scale $\eqref{ell}$ is equal to the value of the instantaneous correlation length at a time $t=\tau$ ($\xi(\tau)=\ell$). Indeed, for $|t|\leq \tau$, the response of the system is very slow and in good approximation the state of the system can be considered frozen at the crossover-time value [ref].\textcolor{black}{[Here a figure which summarize]} Neverthless, we can investigate the dynamics in the off-equilibrium regime $|t|\leq \tau$ by assuming a non trivial rescaling of the time-dependent correlation functions with $\ell$. 
%
% More precisely,  close to the critical point $h\to0$ and in the limit of a slow quench $t_s\to\infty$ such that $t/\tau$ and $q\, \ell$ are kept fixed, 
It is well-known that the time-dependent correlation functions exhibit dynamical scaling behaviour for $h\To 0$ \cite{Sondhi-12}
%1-column
% \begin{equation}\label{KZscaling}
% M(t) \sim \ell^{-d_{\phi}}\; \mathcal{M}(t/\tau), \qquad G_{\perp}(\vec{q},t) \sim \ell^2 \; \mathcal{G}_{\perp}(\vec{q}\, \ell, t/\tau);
% \end{equation}
\begin{subequations}\label{KZscaling}
\begin{align}\label{KZscaling-a}
M(t) \sim \ell^{-d_{\phi}}\; \mathcal{M}(t/\tau)\ ,\\[.25cm]
G_{\perp}(\vec{q},t) \sim \ell^2 \; \mathcal{G}_{\perp}(\vec{q}\, \ell, t/\tau) \ ,\label{KZscaling-b}
\end{align}
\end{subequations}
where $d_{\phi}=(D-2)/2$ is the scaling dimension of the order parameter $M$ and $\mathcal{M}(\cdot)$, $\mathcal{G}_{\perp}(\cdot)$
are generic scaling functions \cite{Henk10}. 
\begin{figure}[ht]
\centering
\includegraphics[width=.485\textwidth]{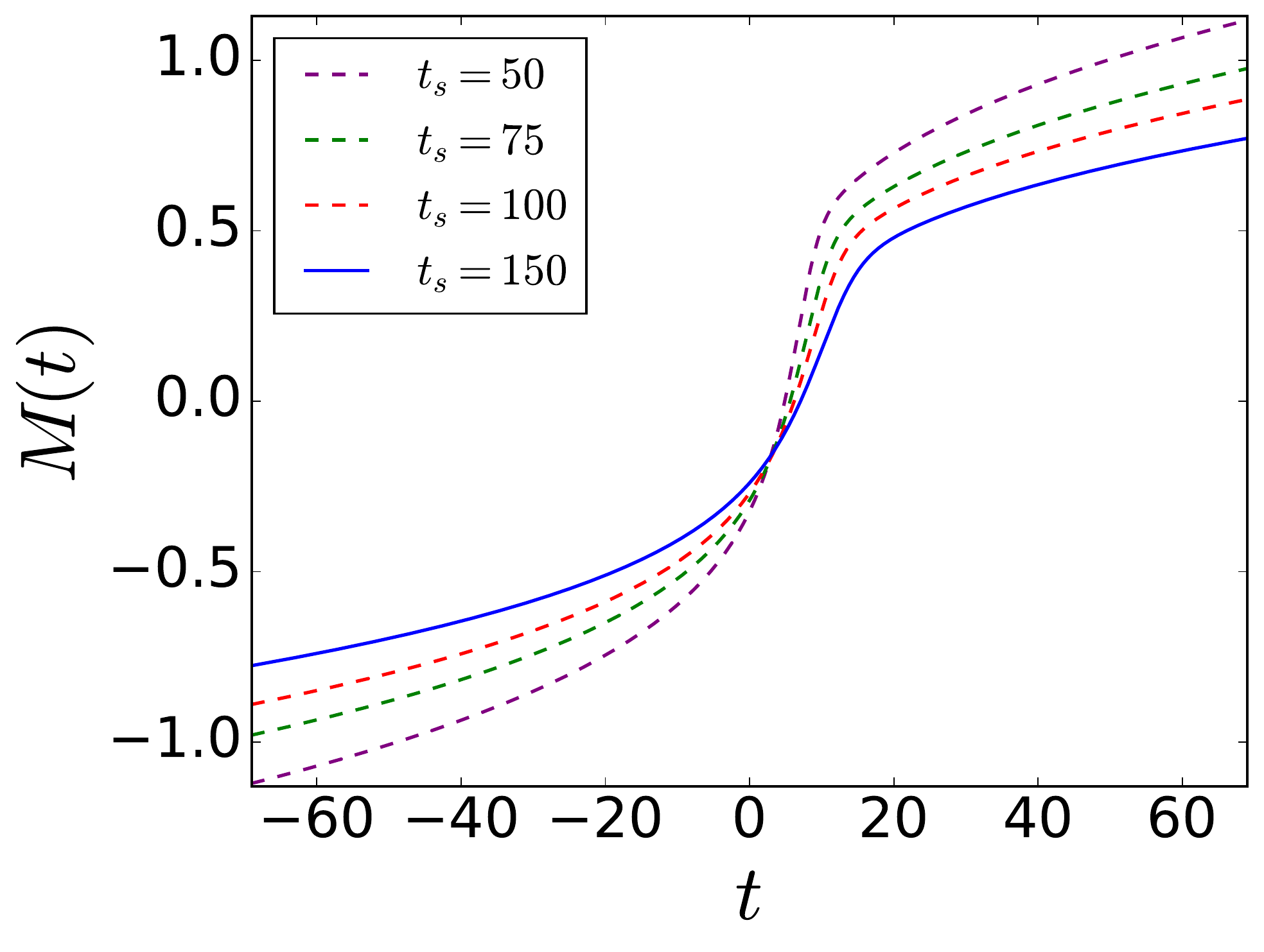} \
\includegraphics[width=.485\textwidth]{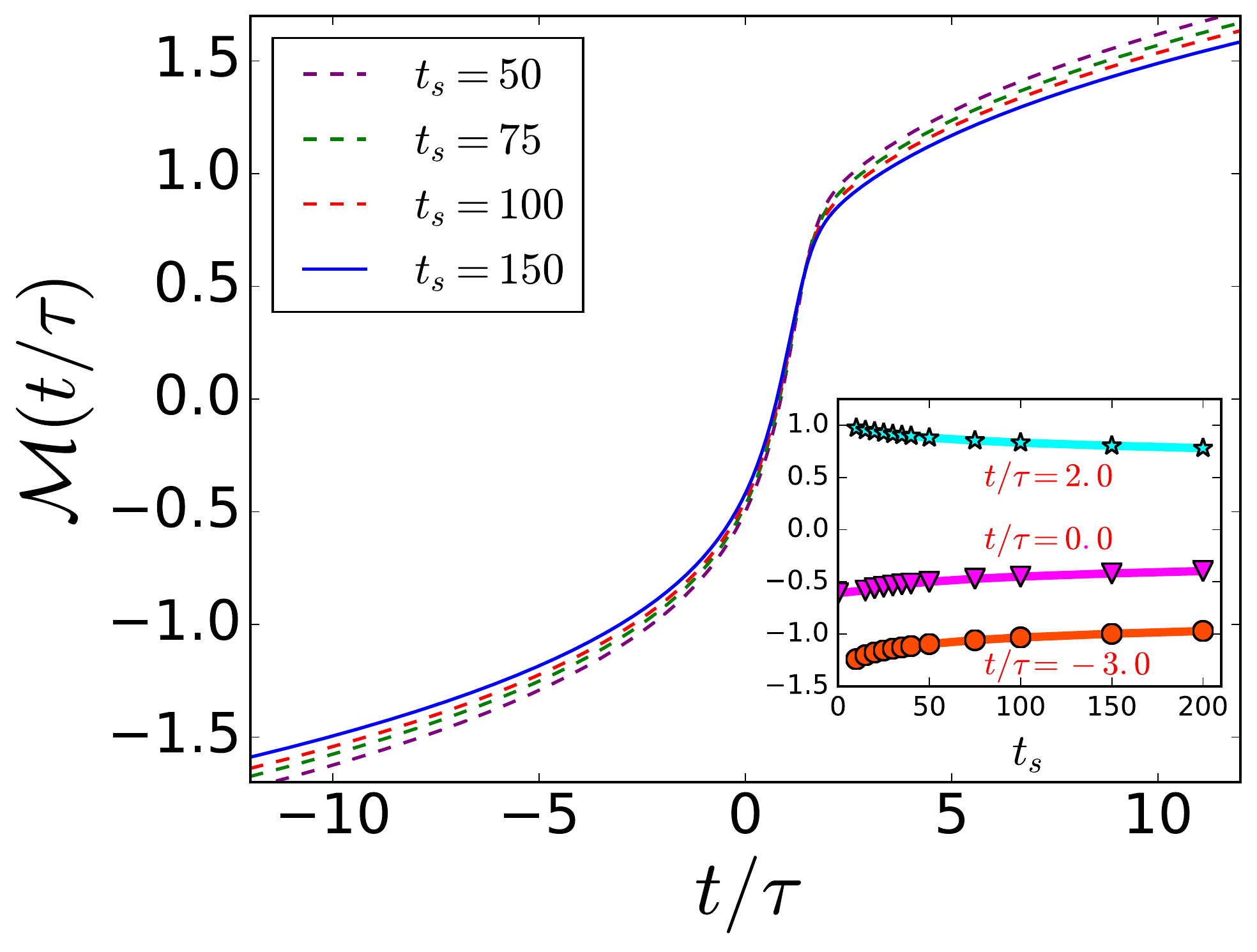}
\caption{\small  Numerical analysis of the dynamical magnetisation in $D=3$ spatial dimensions
%as obtained from the dynamical equations (\ref{dyn-obs}, \ref{dyn-obsG}, \ref{time-eq-state})
and at the critical temperature $T=T_c$.
\underline{Left panel}: magnetisation as a function of time for different quench times $t_s$, see eq~(\ref{KZ-t}).
\underline{Right panel}: data collapse and dynamical scaling function for the magnetisation (compare eq~$\eqref{KZscaling-a}$). \textcolor{black}{The inset shows the convergence of the scaling functions at finite $t_s$ towards the asymptotic regime ($t_s\To \infty$) for three different times.}}
\label{fig:num_M_rc}
\end{figure}

We briefly comment that finite-size scaling can be implemented in this theory as well.
For a system of finite size $L$, we would consider the limit $L,t_s\To\infty$, $t\To 0$ such that $t/\tau$, $q\, \ell$ and $L/\ell$ are fixed. In this 
limit the time-dependent correlators present the scaling relations \cite{Henk10}
%1-column 
%\begin{equation}\label{FSS}
% M(t,L) \sim L^{-d_{\phi}} \; \mathcal{M}(t/\tau, L/\ell), \qquad G_{\perp}(\vec{q},t,L) \sim L^2 \; \mathcal{G}_{\perp}(\vec{q}\, \ell, t/\tau, L/\ell);
% \end{equation}
\begin{subequations}\label{FSS}
\begin{align}
M(t,L) \sim L^{-d_{\phi}} \; \mathcal{M}(t/\tau, L/\ell)\ ,\\[.25cm]
G_{\perp}(\vec{q},t,L) \sim L^2 \; \mathcal{G}_{\perp}(\vec{q}\, \ell, t/\tau, L/\ell)\ ,
\end{align}
\end{subequations}
such that the infinite-volume behaviour of $\eqref{KZscaling}$ is recovered for $L/\ell \To 0$ at fixed $q\, \ell$, $t/\tau$.
Notice also that, by construction, these scaling relations match the equilibrium scaling behaviour for $|t|\To \tau$ (see appendix \ref{equilibrium}).

From a dimensional analysis, it is clear that the effective mass term must scale as \cite{Sondhi-12}
\begin{equation}\label{eq:mscalez2}
m^2(t)\sim \ell^{-2} \, \mathfrak{m}^2(t/\tau)
\end{equation}
with a general scaling function $\mathfrak{m} (\cdot)$.
From eq~(\ref{eq:M-t}), the magnetisation is then given as\footnote{The 
dependence on the initial condition is exponentially suppressed in the scaling limit, see appendix \ref{equilibrium}.}
\begin{subequations}
\begin{equation}\label{scal-M}
\mathcal{M}(\bar{t})=\int_{-\infty}^{\bar{t}} \D u \, u \, \exp\bigg[\!\!-\!\int_{u}^{\bar{t}} \D s \, 
\mathfrak{m}^2(s)\bigg] 
\end{equation}
and from eq~(\ref{eq:GT-t}) the transverse two-point function reads
\begin{equation}\label{scal-G}
\mathcal{G}_{\perp}(\bar{\vec{q}},\bar{t}) =2 \int_{-\infty}^{\bar{t}} \D u \, 
\exp\bigg[-2\int_{u}^{\bar{t}} \D s \, (\bar{\vec{q}}^2+\mathfrak{m}^2(s))\bigg] 
\end{equation}
\end{subequations}
with the rescaled time $\bar{t}=t/\tau$ and momentum $\bar{\vec{q}} = \vec{q}\, \ell$. The time evolution of the scaling functions is thus 
solely determined by the function $\mathfrak{m}^2$ which has to be found from eq~$\eqref{time-eq-state}$\footnote{
We use the shorthand notation $\int_{\bar{\vec{q}}}\equiv \int^{\infty}  \frac{\D \bar{\vec{q}}}{(2\pi)^D}$. Notice that the scaling limit is {\it cut-off independent} since $\Lambda\, \ell \To \infty$ \cite{Sondhi-12}.}
\begin{equation}\label{eq-state-critical}
\mathcal{M}^2(\bar{t}, \mathfrak{m}) = \int_{\bar{\vec{q}}}
(\mathcal{G}_{\perp}(\bar{\vec{q}},\bar{t},0)- \mathcal{G}_{\perp}(\bar{\vec{q}},\bar{t},\mathfrak{m}))
\end{equation}
where the critical thermal coupling constant has been expressed in terms of the critical two-point function \cite{LargeN-rev,Sondhi-12}
\begin{equation}\label{r_c}
r_c=-u \int_q \, G_{\perp}(\vec{q}, t, 0)\ .
\end{equation}
The equation of state $\eqref{eq-state-critical}$ shows that the time evolution of the magnetisation is generated
by a dissipation into the transverse field components. In other words, the deviations from equilibrium of 
the magnetisation and of the transverse correlations compensate each other.

\begin{figure}[ht]
\centering
\includegraphics[width=.475\textwidth]{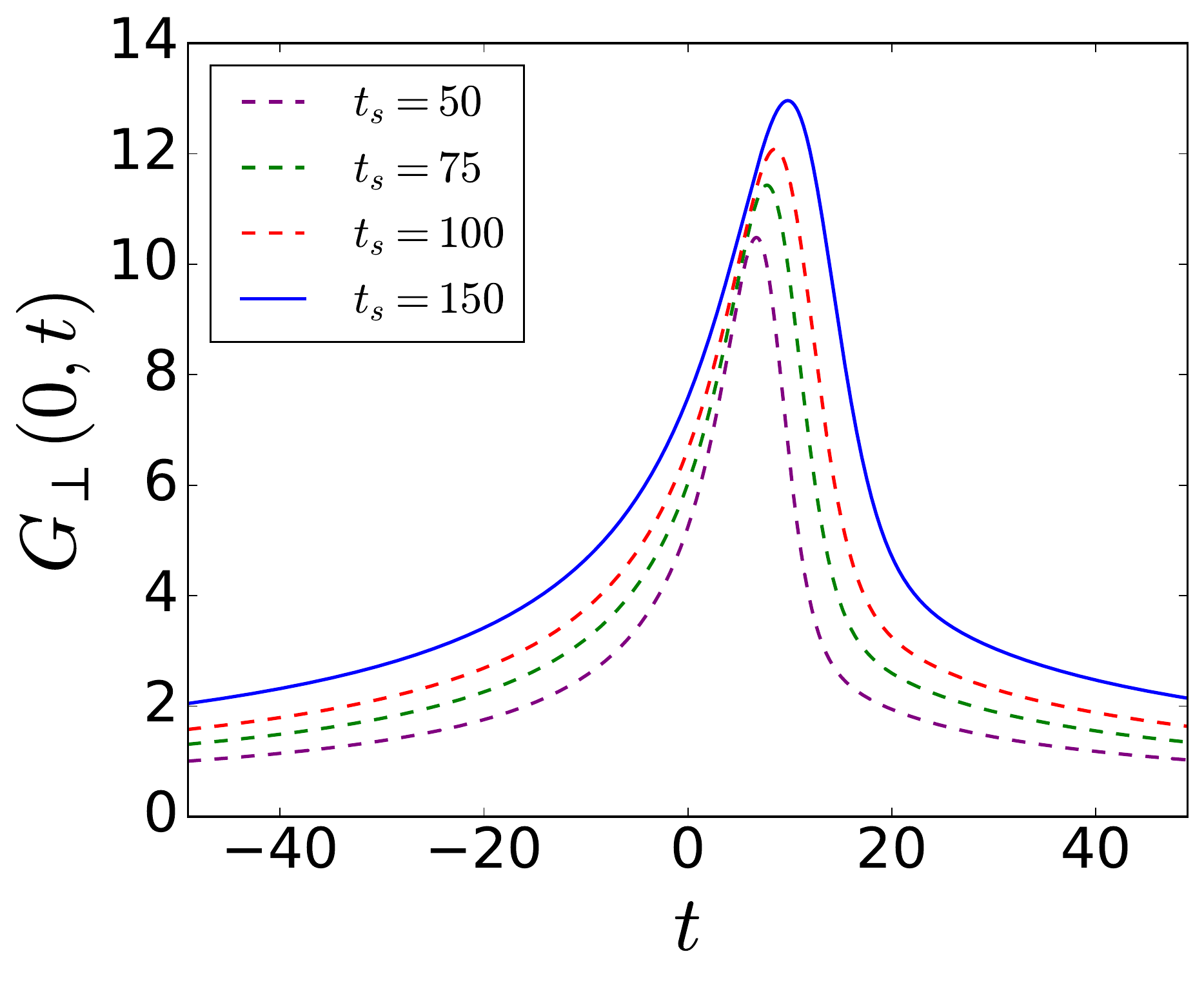} \ 
\includegraphics[width=.49\textwidth]{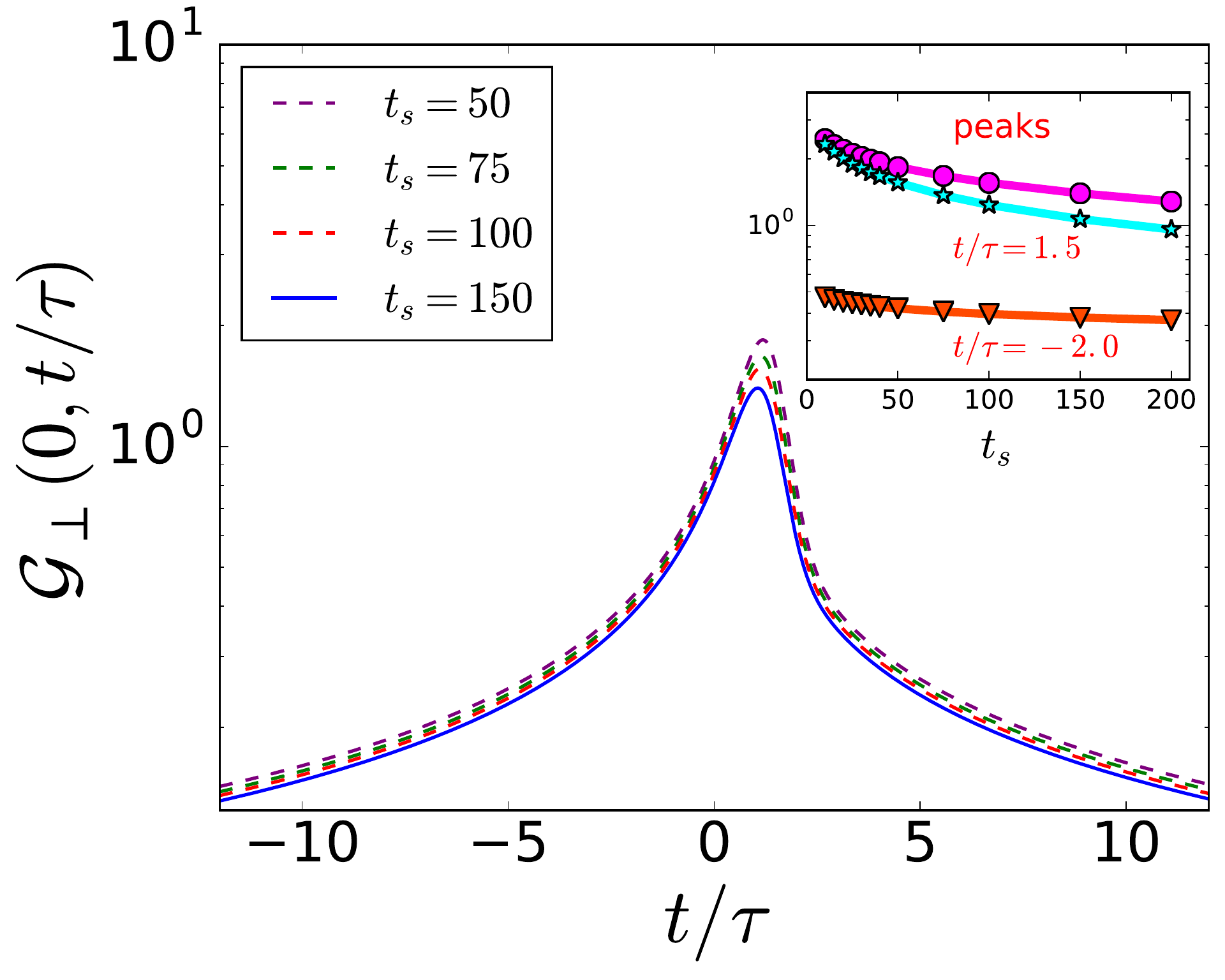}
\caption{\small Numerical analysis of the zero mode correlation function at $T=T_c$ in $D=3$ spatial dimensions. We see the data in the left panel for different
quench times $t_s$, see eq~(\ref{KZ-t}), and the data collapse \blue{(in log scale)} for the dynamical scaling function in the right panel  (compare eq~$\eqref{KZscaling-b}$). \textcolor{black}{The inset shows the convergence of the scaling functions at finite $t_s$ towards the asymptotic regime ($t_s\To \infty$) for three different times} (the label ``peaks'' refers to the convergence of the maximum of the curves).}\label{num_2_rc}
\end{figure}

\begin{figure}[ht]
\centering
\includegraphics[width=.5\textwidth]{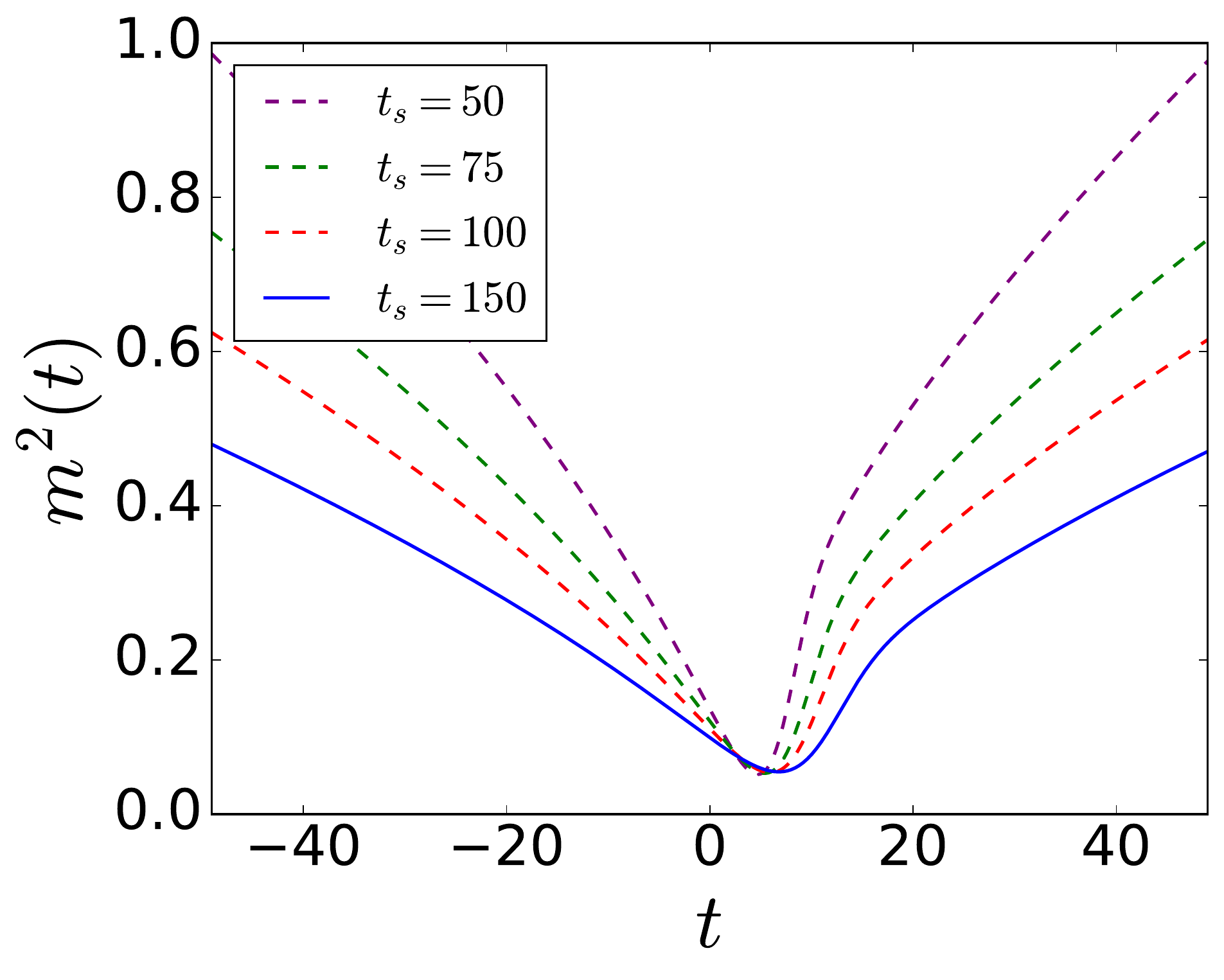} \ 
\includegraphics[width=.475\textwidth]{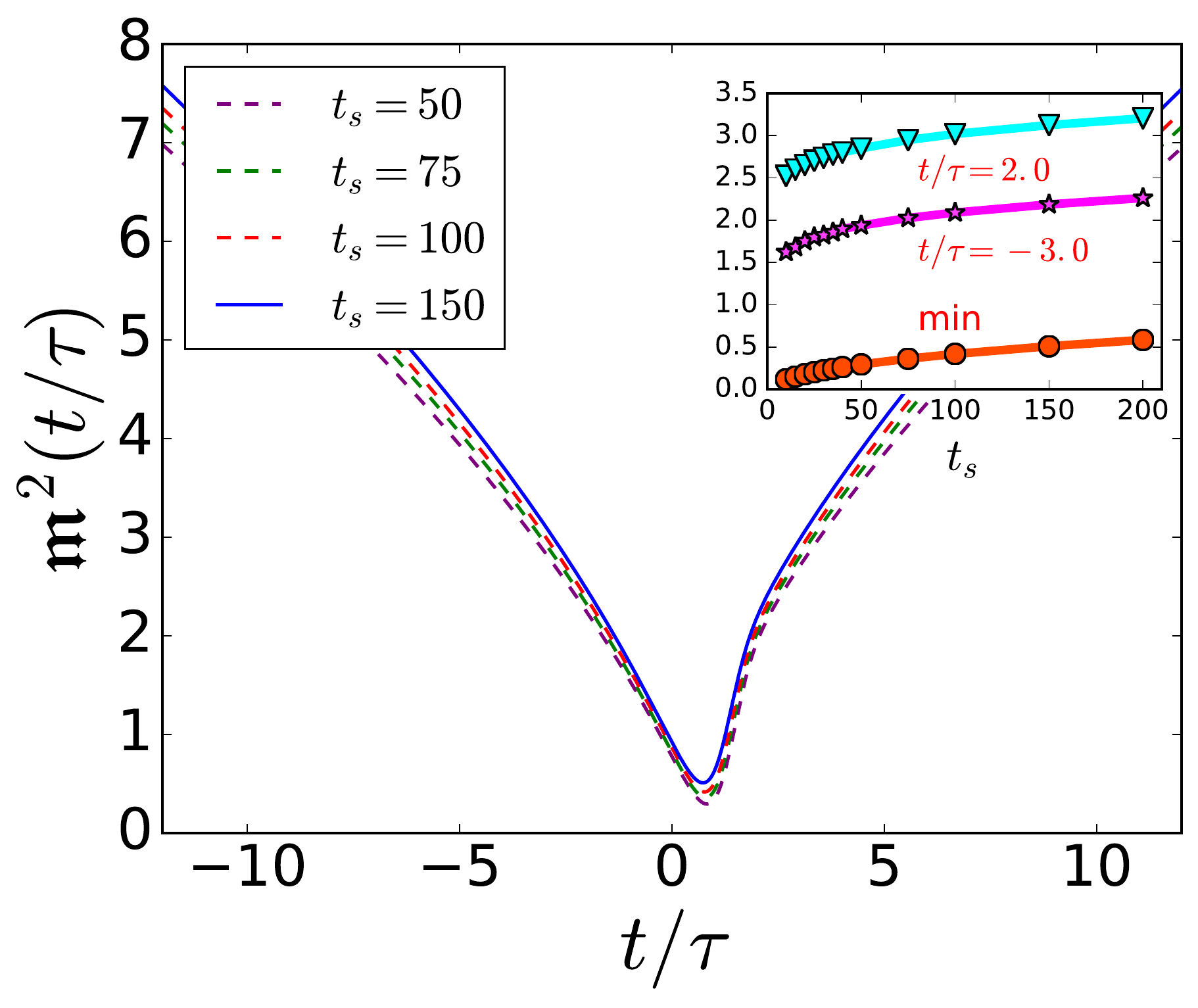}
\caption{\small Numerical analysis of the effective mass at $T=T_c$ in $D=3$ dimensions. We see the data in the left panel for different
quench times $t_s$ and the data collapse for the dynamical scaling function in the right panel (compare eq~$\eqref{eq:mscalez2}$). \blue{The inset shows the convergence of the scaling functions at finite $t_s$ towards the asymptotic regime ($t_s\To \infty$) for three different times}.
}\label{num_2_rc_m}
\end{figure}

In figure \ref{fig:num_M_rc} we show the numerical result for the time evolution of the magnetisation at the critical temperature. 
The numerical analysis confirms our scaling predictions, since clearly, for increasing times $\tau \sim t_s^{\frac{z\nu_h}{1+z \nu_h}}$, 
the data collapse onto a master curve which represents the sought scaling function.
Similar results are obtained for the zero mode correlation function and the for mass term, shown in figure~\ref{num_2_rc} and \ref{num_2_rc_m} respectively.


\subsection{Scaling theory for {\sc fot}s ($T<T_c$)}
\label{ssec:t<tc}

In this section we argue that a {\sc kz}-like theory can also be applied to the magnetic quench performed at 
$T<T_c$.\footnote{In what follows, the scaling theory does not depend on the specific value of the 
temperature $T<T_c$.}
The main obstacle for transferring the {\sc kz} scaling theory to the {\sc fot} below $T_c$ is that 
the system correlation length remains {\it finite}. We shall therefore turn to another length scale,
the so-called \textit{coherence length} or \textit{persistent length} $\xi_h$ \cite{Fisher-82}
which may be defined as the typical size of domains of \blue{(}aligned\blue{)} spins \blue{in the minimum energy configuration}.

\begin{figure}[b]
 \centering
 \includegraphics[width=.95\textwidth]{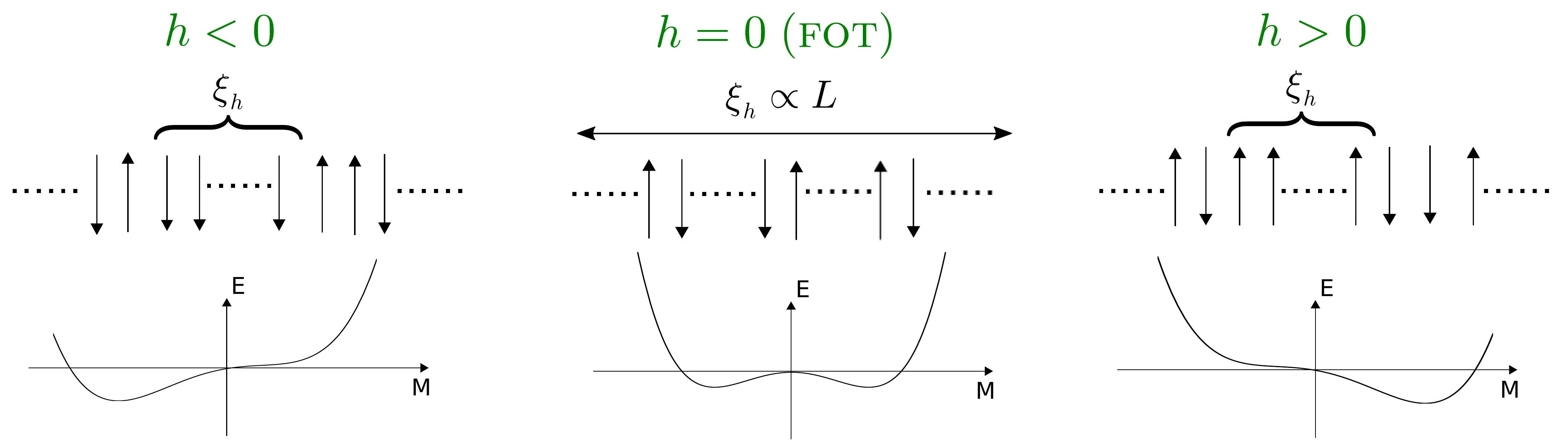}
 \caption{\small Visualisation of the coherence length $\xi_h$ and the associated Ginzburg-Landau functional $E$ as 
 a function of the order parameter $M$ for a generic spin system. 
  The different pictures show the variation of $\xi_H$ and $E$ during the protocol described in eq~$\eqref{eq:quench}$.  At the {\sc fot} $h=0$, the degeneracy of the two vacua associated to the different realisations of the ordered phase leads to a divergence of $\xi_h$.}
 %\textcolor{magenta}{maybe one sentence to point out the difference to the correlation length?}}
 \label{fig:xih}
\end{figure}
For $h\To 0$, the system cannot energetically distinguish between the two ordered phases and long-range order arises 
which leads to an increase of the coherence length. Eventually, this results in a macroscopic coherence length
$\xi_h \propto L$, see figure~\ref{fig:xih}.

% a coherence length which is of the order of the system size $L^D$. 
In order to construct the scaling theory, we need to know the scaling behaviour of the coherence length as a function of the 
magnetic field in analogy to eq~(\ref{eq:corrlength}) and the dynamical exponent $z$. 
It is well-known that the behaviour of $\xi_h$ close to the {\sc fot} can be expressed as a power-law \cite{Nienhuis-75,Fisher-82,Privman-83,Binder}
\begin{equation}
\xi_h(t) \sim |h(t)|^{-1/D} \ \ , \ \  h\To 0 
\end{equation}
from which we can identify the critical exponent
\begin{equation}
 \nu_{h}=1/D \,.
\end{equation}
For the dynamical exponent $z$, a lengthy but straightforward calculation reveals 
\begin{equation}
 z = D \ ,
\end{equation}
which can be qualitatively understood as follows.  Initially, the equilibrium magnetisation is aligned with the initial 
magnetic field $h<0$. As 
this field is driven across the critical value $h_c=0$, the magnetisation has to flip in order to align with the 
final magnetic field $h>0$. Therefore, the vector of the magnetisation has to perform a rotation which needs a 
characteristic time of the order of the
system volume $L^D$\cite{Vicari-off-16}. For further details on how to determine $z$ in the low temperature regime,
we refer to appendix~\ref{fot-dyn}.

We are now able to draw the analogy to eq~(\ref{KZ-t}), i.e. the freeze-out condition reveals
\begin{equation}\label{scales-fot}
\tau_{\text{\sc fot}} = \sqrt{t_s}, \qquad \ell_{\text{\sc fot}} =t_s^{1/2D}.
\end{equation}
We notice that the freeze-out time $\tau_{\text{\sc fot}}$ coincides with the \textit{coercive time} of the 
model \cite{Dhar-92} that is the typical time scale after which a ferromagnet reacts to an inversion of the 
external magnetic field.

\begin{figure}[ht]
\centering
\includegraphics[width=0.475\textwidth]{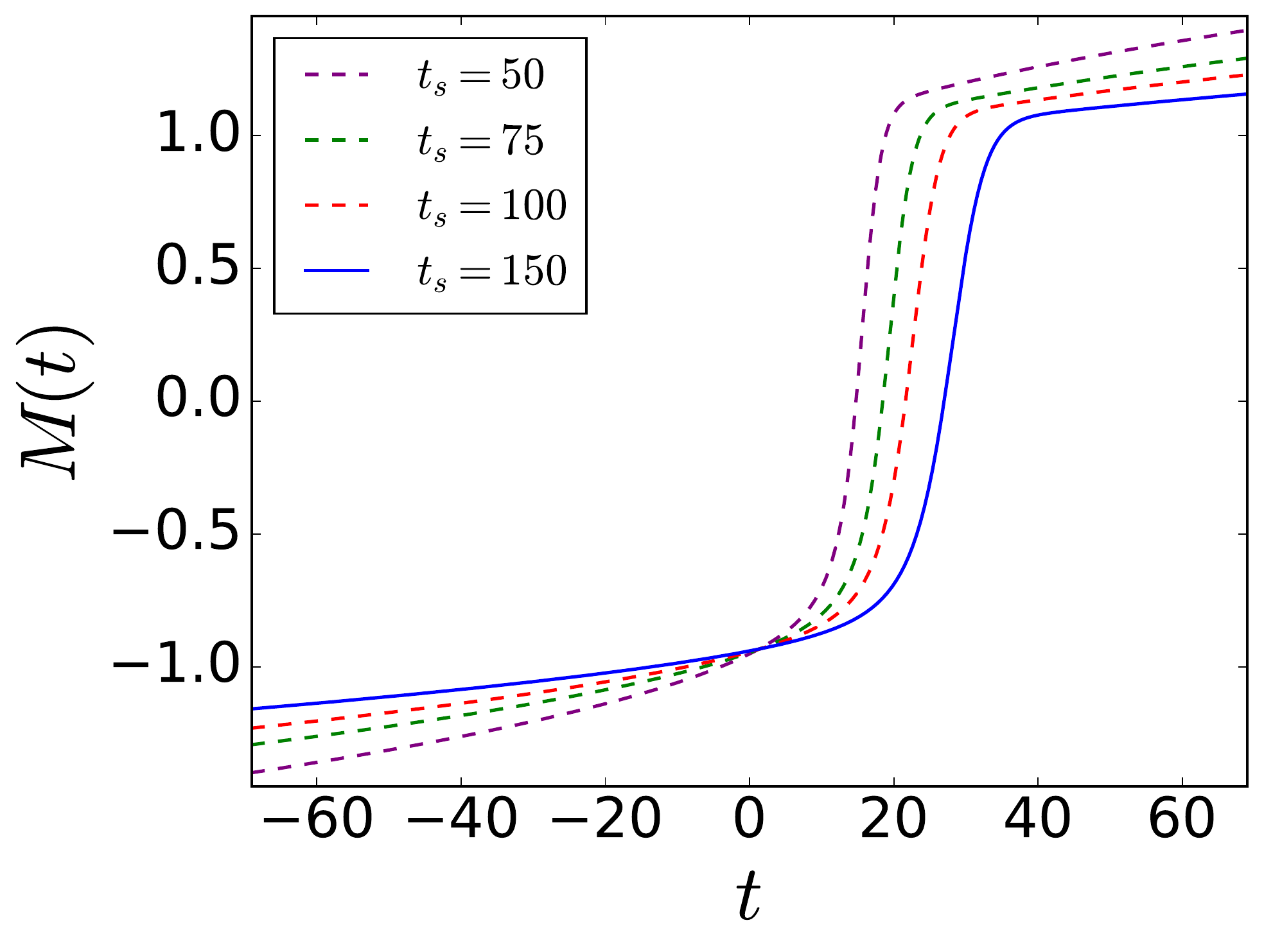} \ 
\includegraphics[width=0.475\textwidth]{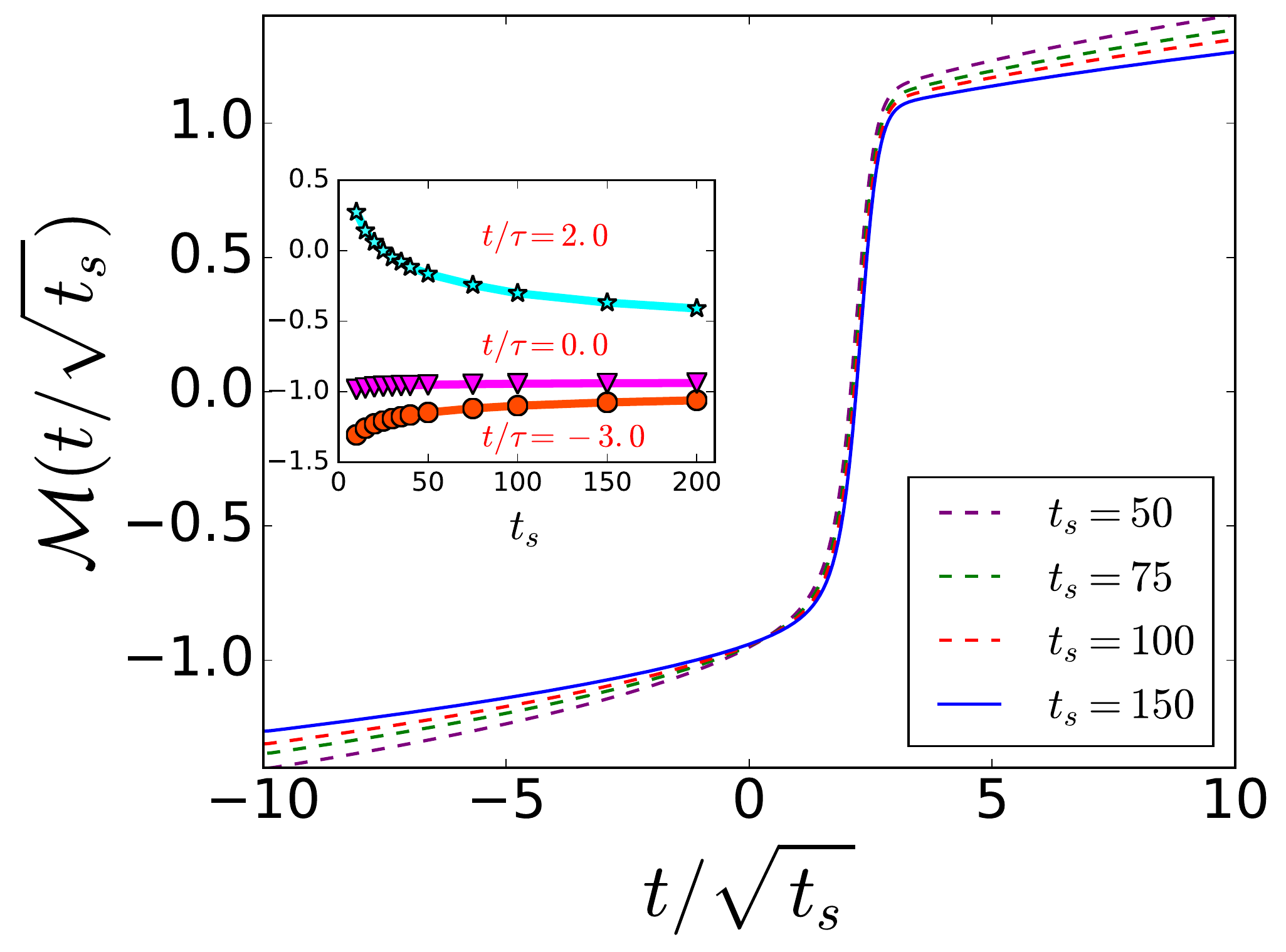}
\caption{\small Numerical analysis of the dynamical magnetisation below $T_c$ ($r=-1$) in $D=3$ spatial dimensions. \underline{Left panel}: dynamical magnetisation 
as a function of time for different quench time scales $t_s$. \underline{Right panel}: data collapse and dynamical 
scaling function for the magnetisation (compare eq~$\eqref{scaling-fot}$). \blue{The inset shows the convergence of the scaling functions at finite $t_s$ towards the asymptotic regime ($t_s\To \infty$) for three different times}.}
\label{num_1_low}
\end{figure}

Extending this analogy further, we assume that the time-dependent magnetisation of the system 
in the vicinity of the transition point $h\to0$ shows dynamical
scaling behaviour in line with eq~(\ref{KZscaling}) in the limit $t_s\to\infty$ (with $t/\tau_{\text{\sc fot}}$ fixed)
%
%1-column
% \begin{equation}\label{scaling-fot}
% M(t) \sim \ell_{\text{\sc fot}}^{-d_{\phi}}\; \mathcal{M}(t/\tau_{\text{\sc fot}}), \qquad G_{\perp}(\vec{q},t) \sim \ell_{\text{\sc fot}}^2 \; \mathcal{G}_{\perp}(\vec{q}\, \ell_{\text{\sc fot}}, t/\tau_{\text{\sc fot}});
% \end{equation}
%\begin{subequations}
 \begin{equation}
 \label{scaling-fot}
%\begin{align}
M(t) \sim \ell_{\text{\sc fot}}^{-d_{\phi}}\; \mathcal{M}(t/\tau_{\text{\sc fot}})\ ,\\
\end{equation}
%G_{\perp}(\vec{q},t) &\sim \ell_{\text{\sc fot}}^D \; \mathcal{G}_{\perp}(\vec{q}\, \ell_{\text{\sc fot}}, t/\tau_{\text{\sc fot}})\ .
%\end{align}
%\end{subequations}
where the scaling dimension of the order parameter is known to be $d_{\phi}=0$ \cite{Fisher-82}. 
The numerical result for the dynamical magnetisation below the critical temperature is shown in figure \ref{num_1_low}
and explicitly verifies our scaling prediction (\ref{scaling-fot}).

\begin{figure}[ht]
\centering
\includegraphics[width=.485\textwidth]{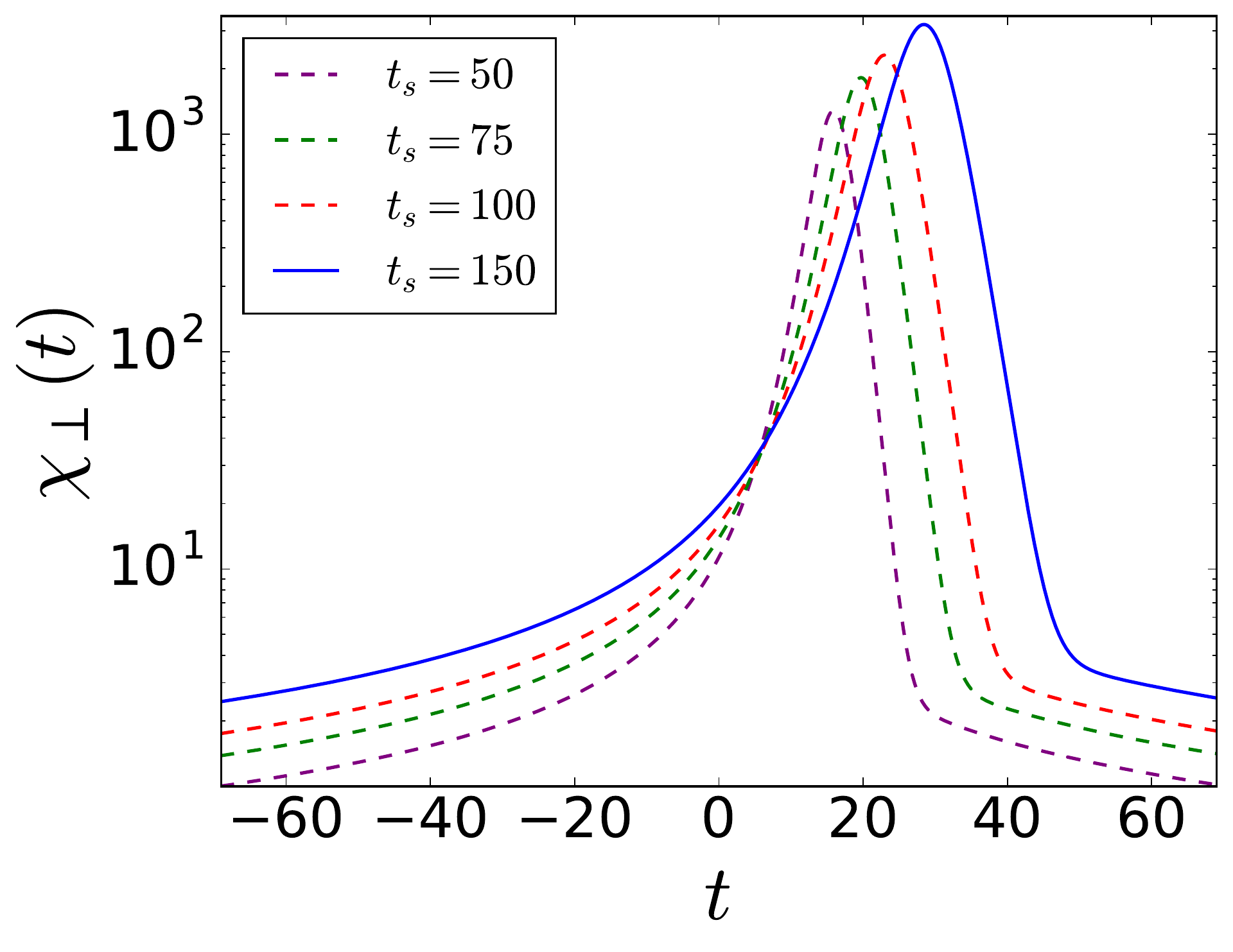} \ 
\includegraphics[width=.485\textwidth]{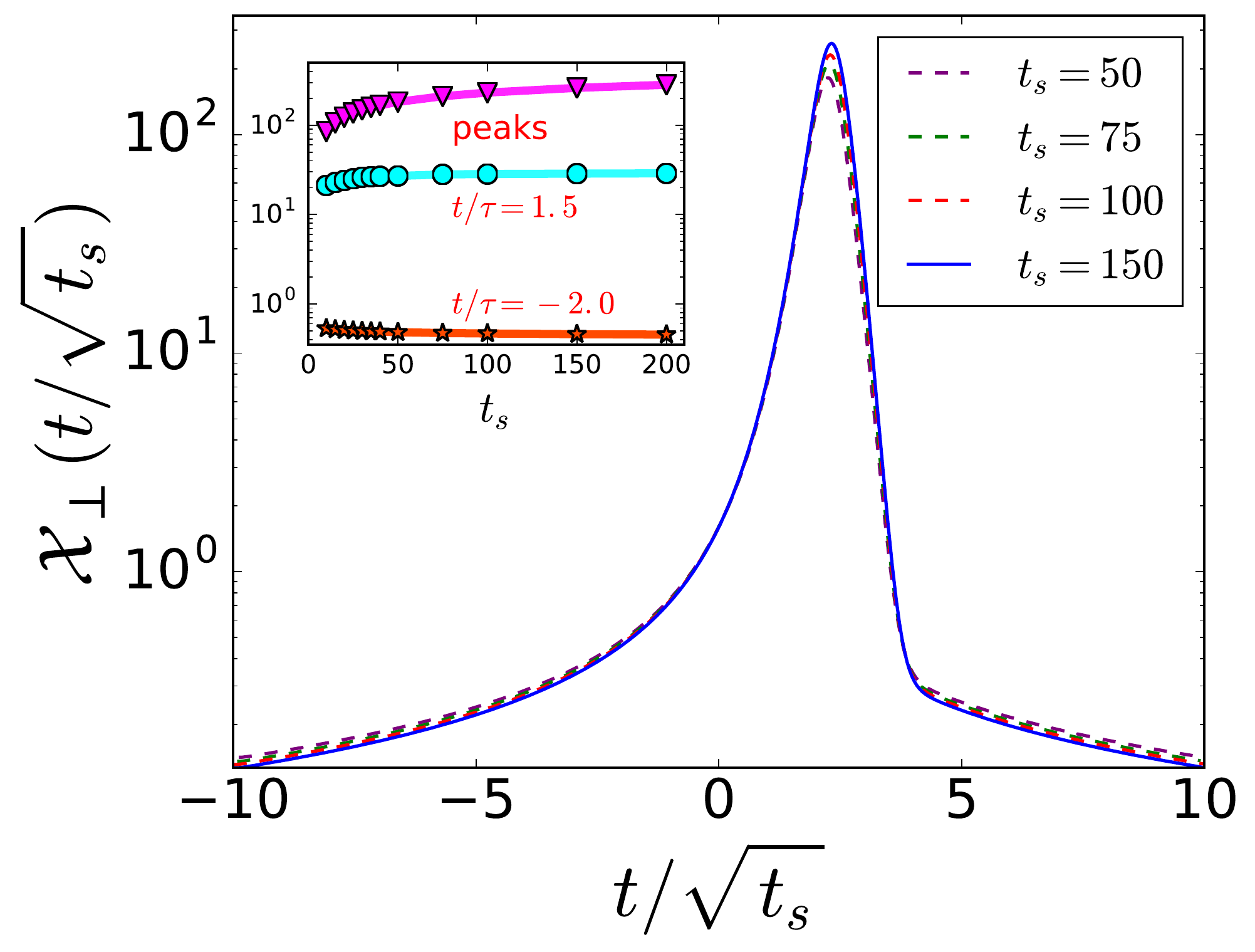}
\caption{\small Numerical analysis of the transverse susceptibility below $T_c$ ($r=-1$) in $D=3$ spatial dimensions. 
In the left panel, the data for different
quench times $t_s$ is shown \blue{in log scale} and the right panel shows the data collapse for the dynamical scaling function (compare eq~$\eqref{scal-G-low}$). \blue{The inset shows the convergence of the scaling functions at finite $t_s$ towards the asymptotic regime ($t_s\To \infty$) for three different times}.}\label{num_2_low}
\end{figure}

\begin{figure}[ht]
\centering
\includegraphics[width=.5\textwidth]{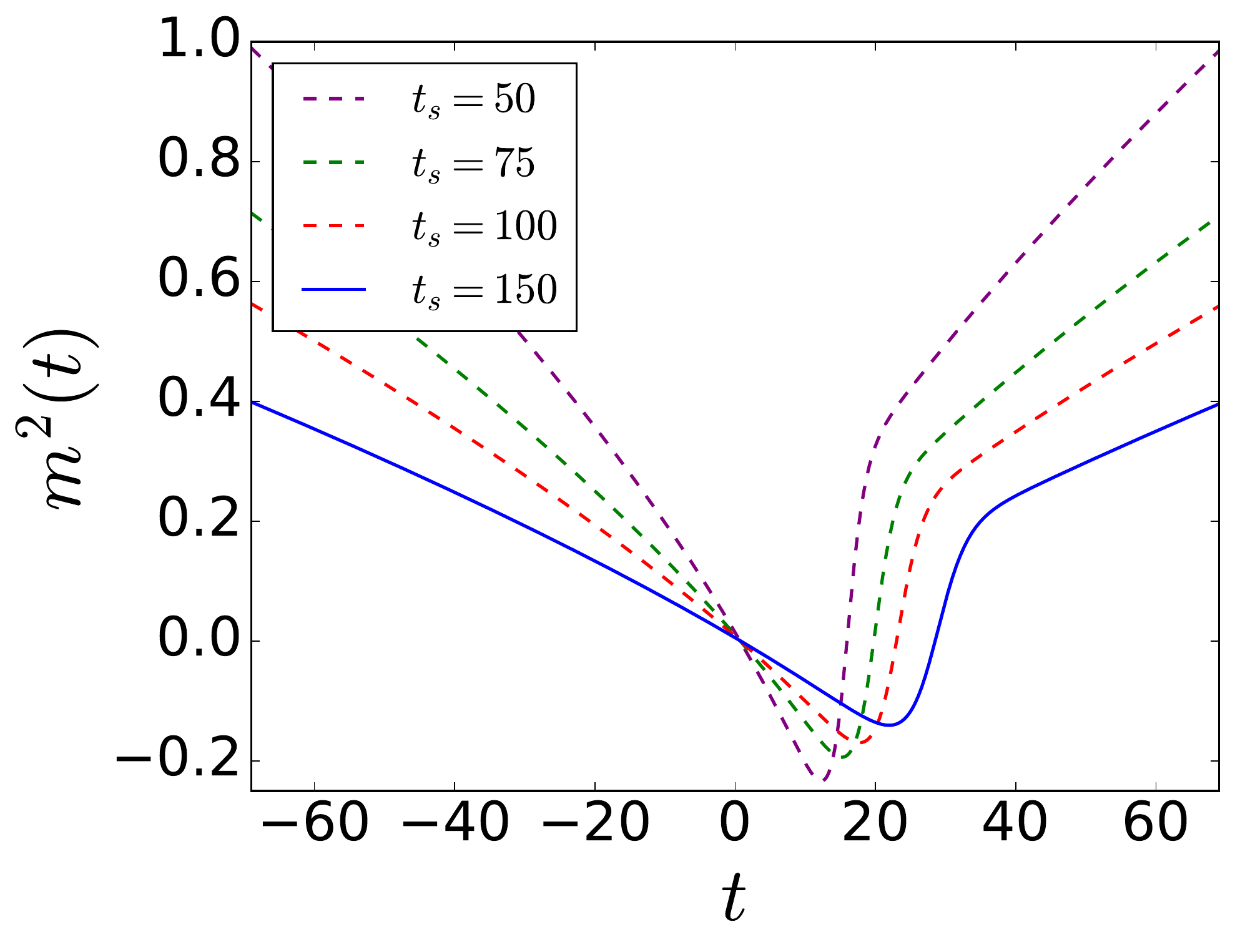} \ 
\includegraphics[width=.45\textwidth]{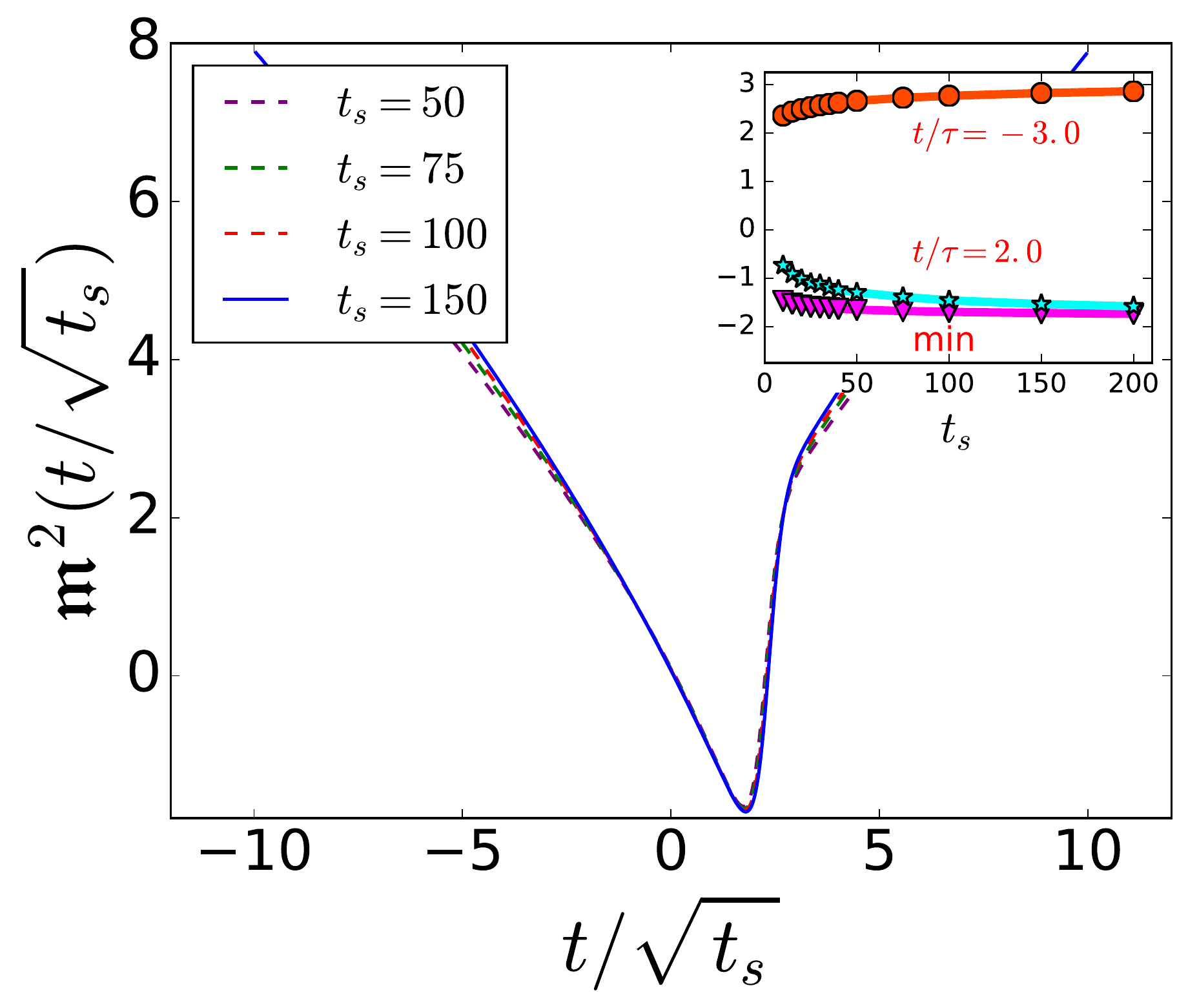}
\caption{\small Numerical analysis of the effective mass below $T_c$ ($r=-1$). We see the data in the left panel for different
quench times $t_s$ and the data collapse for the dynamical scaling function in the right panel (compare eq~$\eqref{eq:mfot}$). Notice that 
the effective mass term can also take negative values below $T_c$, see also \cite{Dhar-92}. This is not surprising in the presence of a broken symmetry and refers to the formation and the propagation of massless modes who connects degenerate vacua, see e.g. \cite{ZinnJustin}.
\blue{The inset shows the convergence of the scaling functions at finite $t_s$ towards the asymptotic regime ($t_s\To \infty$) for three different times}.
}\label{num_3_low}
\end{figure}

We describe our model in the \textit{spin-wave approximation} \cite{SpinWave1,SpinWave2}
which states that at low-temperatures $T<T_c$ it is sufficient to study long-range excitations, i.e. 
only the degrees of freedom with $|\vec{q}| < q^{\ast}$ turn out to be relevant for the off-equilibrium dynamics. The
boundary value $q^{\ast}$ which separates the short-distance fluctuations $|\vec{q}| > q^{\ast}$ from the low-energy 
modes $|\vec{q}| < q^{\ast}$ can be estimated as
$q^{\ast}\propto t_s^{-1/4}$, see appendix~\ref{spin-wave}. In the scaling limit  $h\To 0$, $t_s\To \infty$ keeping 
$q\, \ell_{\text{\sc fot}}$ fixed we notice that 
\begin{equation}
|\vec{q}|\, \ell_{\text{\sc fot}} < q^{\ast} \, \ell_{\text{\sc fot}} \propto t_s^{\frac{2-D}{4D}} \To 0 \ ,
\end{equation}
which implies that the zero-momentum contribution alone provides a good description of the off-equilibrium behaviour arising during 
the quench in the asymptotic limit. We introduce therefore the transverse susceptibility $\chi_{\perp}$ that obeys
the scaling relation
\begin{equation}\label{chi-low}
\chi_{\perp}(t)\equiv G_{\perp}(\vec{0},t) \sim \ell_{\text{\sc fot}}^D \,  \mathcal{X}_{\perp}(\bar{t})\ ,
\end{equation}
%
%or equivalently,
%\begin{equation}
%G_{\perp}(q,t) \sim (q^*)^{-2} {\cal G}_{\perp}(\vec{0}, t/\tau)
%\end{equation}
as shown in figure \ref{num_2_low}.
Notice that, by construction, the off-equilibrium scaling behaviours (\ref{scaling-fot},\ref{chi-low}) match again the equilibrium scaling 
when $|t|\To \tau_{\text{\sc fot}}$ (see appendix \ref{equilibrium}).
For the effective mass we write the analogous scaling behaviour to eq~(\ref{eq:mscalez2})
\begin{equation}
m^2(t) \sim \ell_{\text{\sc fot}}^{-D} \; \mathfrak{m}^2(t/\tau_{\text{\sc fot}})
\label{eq:mfot}
\end{equation}
due to the presence of a small magnetic field with scaling dimension $d_h=D$.\footnote{From
eq~$\eqref{dyn-obs}$ we may write $m^2(t) =M(t)^{-1}\left(h(t) +\frac{\D}{\D t}M(t)\right)$ from which we 
conclude eq~(\ref{eq:mfot}).} The numerical result for the effective mass is shown in figure \ref{num_3_low}.

The time-dependent magnetisation satisfies
the (trivial) scaling relation in eq~$\eqref{scaling-fot}$ with the scaling function
\begin{equation}\label{scal-M-low}
M(t)\equiv \mathcal{M}(\bar{t})=\int_{-\infty}^{\bar{t}}  \D s  \, s \, \exp\bigg[-\int_{s}^{\bar{t}}
\!\D u \, \mathfrak{m}^2(u)\bigg],
\end{equation}
while the scaling function of the transverse susceptibility reads
\begin{equation}\label{scal-G-low}
\mathcal{X}_{\perp}(\bar{t}) =2 \int_{-\infty}^{\bar{t}} \D s \, \exp\bigg[-2\int_{s}^{\bar{t}} \D u 
\, \mathfrak{m}^2(u)\bigg] \ ,
\end{equation}
where now $\bar{t}=t/\tau_{\text{\sc fot}}$ was redefined. In the spin-wave approximation we introduce the quantity
\begin{equation}
\s(t)=\int_q \, G_{\perp}(|\vec{q}|<q^{\ast}, t) \sim {\cal S}(t/\tau_{\text{\sc fot}})
\end{equation}
with a trivial scaling relation that follows from eq~$\eqref{chi-low}$.

%In the spin-wave approximation we can compute the integral
%%1-column
%% \begin{align}
%% \s(t)&=\int_q \, G_{\perp}(|\vec{q}|<q^{\ast}, t) \\
%% &= \Omega_D \, \int_{t_0}^t dt'\, \sqrt{\frac{\pi}{2}}\frac{\text{Erf}(q^{\ast}\sqrt{2}\sqrt{t-t'})}{2\sqrt{t-t'}}\, 
%% \exp\bigg[-2\int_{t'}^t dt''\, m^2(t'')\bigg]
%% \end{align} 
%%
%
%\begin{align}\nonumber
%\s(t)&=\int_q \, G_{\perp}(|\vec{q}|<q^{\ast}, t) \\ \nonumber
%&= \frac{\Omega_D}{(2\pi)^D} \int_{t_0}^t \D u\, \frac{\Gamma(\frac{D}{2})-\Gamma\left[\frac{D}{2}, 2(q^*)^2\left(t-u\right)\right]}{[2(t-u)^{D/2}]}\,  \exp\bigg[-2\int_{u}^t \D s\, m^2(s)\bigg]
%\end{align} 
%%
%where $\Omega_D=2^{1-D} \pi^{-D/2}/\Gamma(D/2)$ is the surface of the $D$-dimensional unit sphere. The scaling limit of $\s(t)\sim \mathcal{S}(\bar{t})$ reads
%%1-column
%% \begin{equation}
%% \s(t)\sim \mathcal{S}(\bar{t})=  \Omega_D \, \int_{-\infty}^{\bar{t}} \D s\, \sqrt{\frac{\pi}{2}}
%% \frac{1/2}{( \bar{t}-s)^{1/2}}\, \exp\bigg[-2\int_{s}^{\bar{t}} \D u \, \mathfrak{m}^2(u)\bigg].
%% \end{equation}
%\begin{equation}
%\mathcal{S}(\bar{t})=  \frac{\Omega_D}{2} \sqrt{\frac{\pi}{2}}\int_{-\infty}^{\bar{t}} \frac{\D s}{( \bar{t}-s)^{1/2}}\,
%\exp\bigg[-2\int_{s}^{\bar{t}} \D u \, \mathfrak{m}^2(u)\bigg] \ .
%\end{equation}
%\\
Notice that once again the scaling functions above depend implicitly on $\mathfrak{m}^2$
via the equation of state $\eqref{time-eq-state}$ which in the scaling limit and for $T<T_c$ reads
\begin{equation}\label{eq-state-below}
M_0^2 = {\cal M}^2(\bar{t}, \mathfrak{m}) +  \mathcal{S}(\bar{t},\mathfrak{m})-\mathcal{S}(\bar{t},0)  \ ,
\end{equation}
where we have expressed the thermal coupling $r<r_c$ as\footnote{This relation can be easily deduced from eq~$\eqref{time-eq-state}$  considering a system prepared in equilibrium without external fields $h=0$. In this case $m^2=0$ and the magnetisation is equal to $M_0=\sqrt{(r_c-r)/u}$.}
\begin{equation}
r= -u \left( M_0^2  + \mathscr{S}(t, 0) \right).
\end{equation}
Eq~$\eqref{eq-state-below}$ states that the magnetisation deviates from its equilibrium value $|M_0|= \sqrt{(r_c-r)/u}$ 
(see e.g. \cite{ZinnJustin}) by dissipating in the transverse modes.
The magnetisation may be viewed as a $n$-vector $M_{a}(t)\equiv \braket{\phi_{a}(\vec{x},t)}$ whose 
longitudinal component $M(t)$ in eq~$\eqref{eq:M-t}$ is coupled to the other components 
$M_a(t)$, $a>1$, through the transverse correlation function. For weak magnetic fields, we may interpret the magnetisation
as a $n$-vector of fixed length $|M_0|$ whose longitudinal component $M$ is decreased in favour of the transverse modes. 
The dynamical behaviour across the transition point is then nothing but a rotation of this vector. Moreover, this 
kind of dynamics is compatible with the $O(n-1)$ symmetry, because the $n-1$ transverse planes
are equally likely to contain the vector magnetisation at any time. One may also verify that the definition
of the scales $\eqref{scales-fot}$ is the only compatible with the dynamics $z=D$ that preserves the 
equilibrium limit at $|t|\To \tau_{\text{\sc fot}}$.
\begin{figure}[ht]
\centering
\includegraphics[width=.475\textwidth]{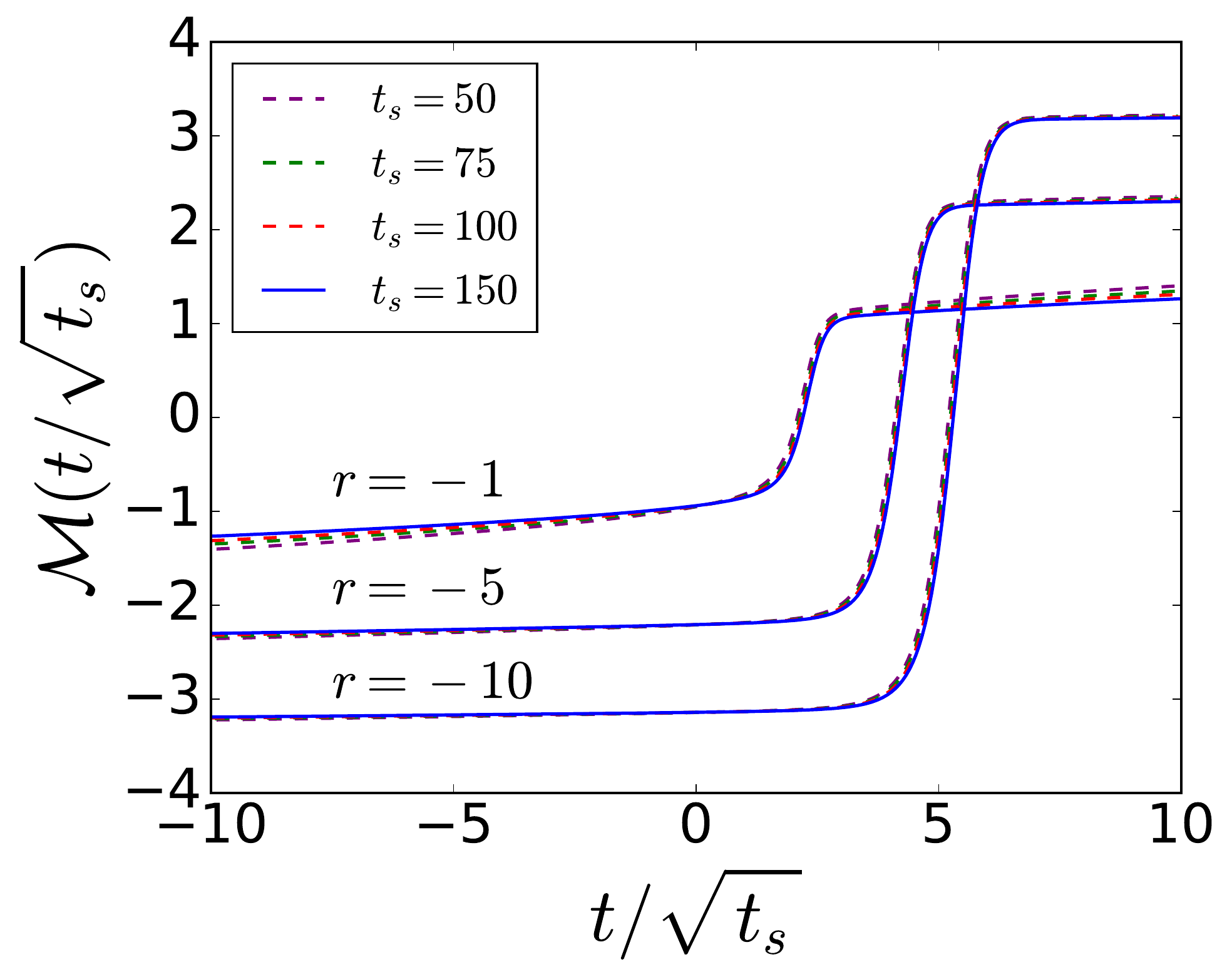} \
\includegraphics[width=.475\textwidth]{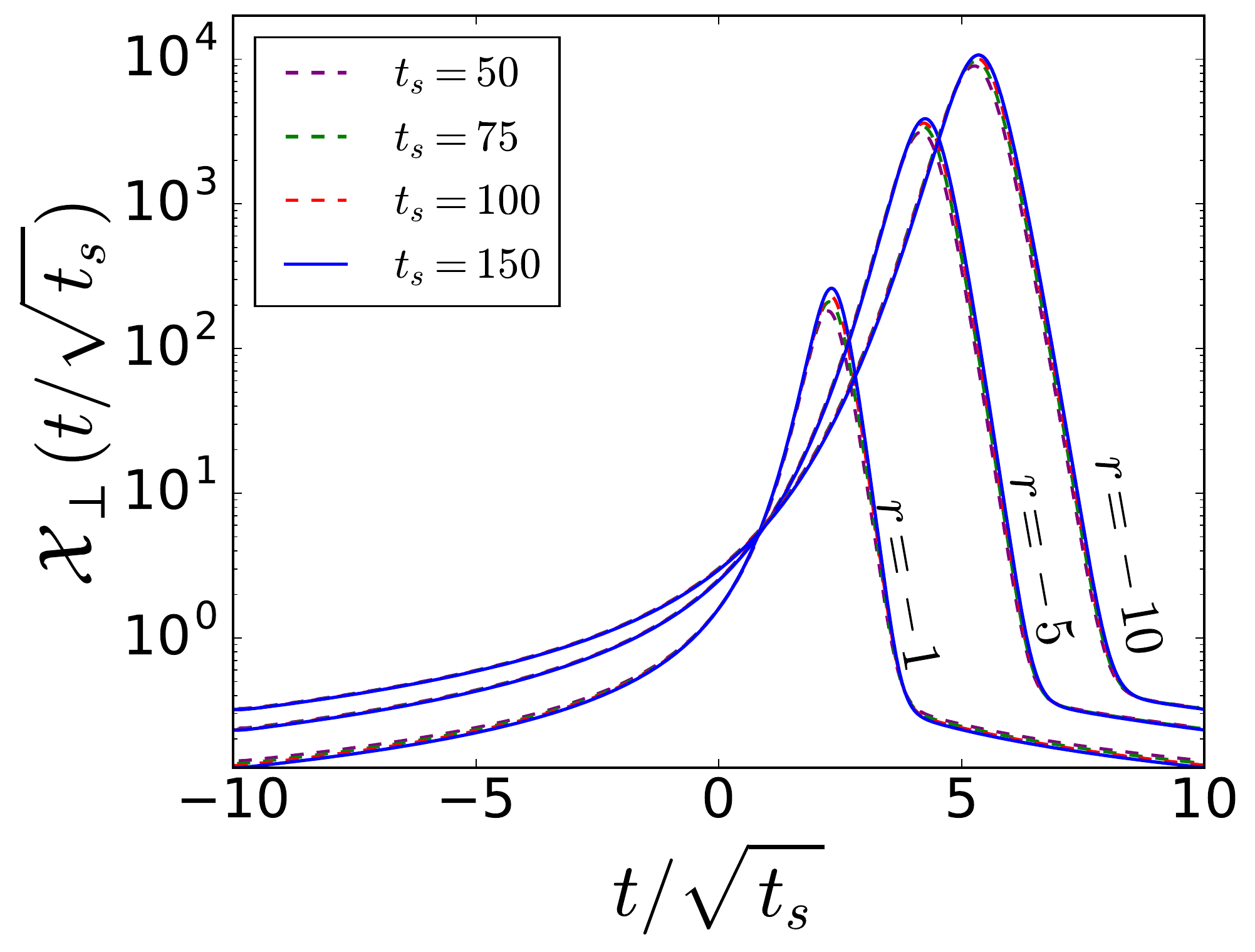}
\caption{\small Off-equilibrium scaling behaviour of the magnetisation (left panel) and of the transverse susceptibility (right panel) for different
values of the temperature ($r=-1,\, -5,\,-10$) below the critical value in $D=3$ spatial dimensions. The data collapse is observed for each value of the temperature confirming 
that the scaling behaviour is not modified by the specific value of $T<T_c$ considered.}\label{num_low_temp}
\end{figure}

In figure \ref{num_low_temp} we numerically investigate the off-equilibrium scaling across the {\sc fot} for different 
values of the temperature $T<T_c$.
As expected from phase kinetic arguments \cite{Bray94}, the scaling theory does not depend on the specific value of the temperature considered.

%########################################################
%########Hysteresis in the round-trip protocol###########
%########################################################
\section{Hysteresis in the round-trip protocol}
\label{sec:hys}

%\begin{figure}
% \centering
%   \includegraphics[width=.4\textwidth]{magn.png}\qquad \includegraphics[width=.4\textwidth]{magn2.png}\\
%  \hspace{1cm} ({\bf a})\hspace{6.75cm} ({\bf b})\hspace{1cm}
%\caption{\small Representation of the rescaled hysteresis loop area ${\cal A}$ in the adiabatic-impulse approximation: we approximate the value of the rescaled magnetisation ${\cal M}$ as frozen in the off-equilibrium regime. This 
%picture may be interpreted as the formation of a metastable state in the off-equilibrium regime that suddenly collapses to the new equilibrium value when the adiabatic regime is again approached.
%({\bf a}) The rescaled hysteresis loop area at $T=T_c$ where the equilibrium magnetisation reads ${\cal M}(u) \sim M_0 \, |u|^{1/5}$, $M_0$ a constant, in three spatial dimensions (see appendix \ref{equilibriumA}). ({\bf b}) The rescaled hysteresis loop area for $T<T_c$ where the equilibrium magnetisation is assumed to be a constant $M_0$ (see appendix \ref{equilibriumB}).}\label{fig:hyst}
%\end{figure}

In this section, we consider a {round-trip protocol} $\gamma(h)$ in which the magnetic field $\eqref{eq:quench}$ 
is varied from an initial value $h(t_i) < 0$ to $h(t_f) > 0$ across the transition point $h_c=0$ at $t=0$
and back in the reversed manner. By integrating the curve described by the magnetisation in time $\eqref{eq:M-t}$ over $\gamma(h)$ 
we obtain the \textit{hysteresis loop area} $A$
%1-column
% \begin{equation}
% A\equiv\oint_{\gamma(h)} \D t \; M(t) = 2\int_{t_0}^{t_f} \D t \, \int_{t_0}^t \D t' \, h(t') \, \cosh\bigg[
% \int_{t'}^t \D t'' \, m^2(t') \bigg]
% \end{equation}
\begin{align}
A\equiv&\oint_{\gamma(h)} \D t \; M(t) =\ 2\int_{t_i}^{t_f} \D t \, \int_{t_i}^t \D u \, h(u) \cosh\bigg[
\int_{u}^t \D s \, m^2(s) \bigg]
\end{align}
which is a quantifier of the deviation from equilibrium during the process \cite{Vicari-off-16,KZ-hyst}: the larger
the deviations from equilibrium are, the larger is the value of $A$, while for a system in equilibrium $A=0$ 
since the magnetisation only depends on the instantaneous value of the external field.
The hysteresis loop area $A$ is related to the magnetic energy $W$ dissipated by the system during the
round trip. For a linear quench $\eqref{eq:quench}$ we have\footnote{This picture is 
compatible with a quasi adiabatic quench where $t_s\to\infty$ and therefore $W=0$ since 
the system will not fall out of equilibrium.}
\begin{equation}
W\equiv \oint_{\gamma(h)} \D h \; M(h)= \frac{A}{t_s}\; .
\end{equation}
%
% which leads to the interpretation of the off-equilibrium behaviour as an energy cost. 
For further considerations, we shall work in the scaling limit $t_s\To\infty$, $h\To 0$ at $\bar{t}$, $|\bar{\vec{q}}|$ fixed (where we refer to the definitions 
$\eqref{ell}$ for $T = T_c$ and $\eqref{scales-fot}$ $T <T_c$ respectively). 

\begin{figure}[ht]
\centering
\includegraphics[width=.475\textwidth]{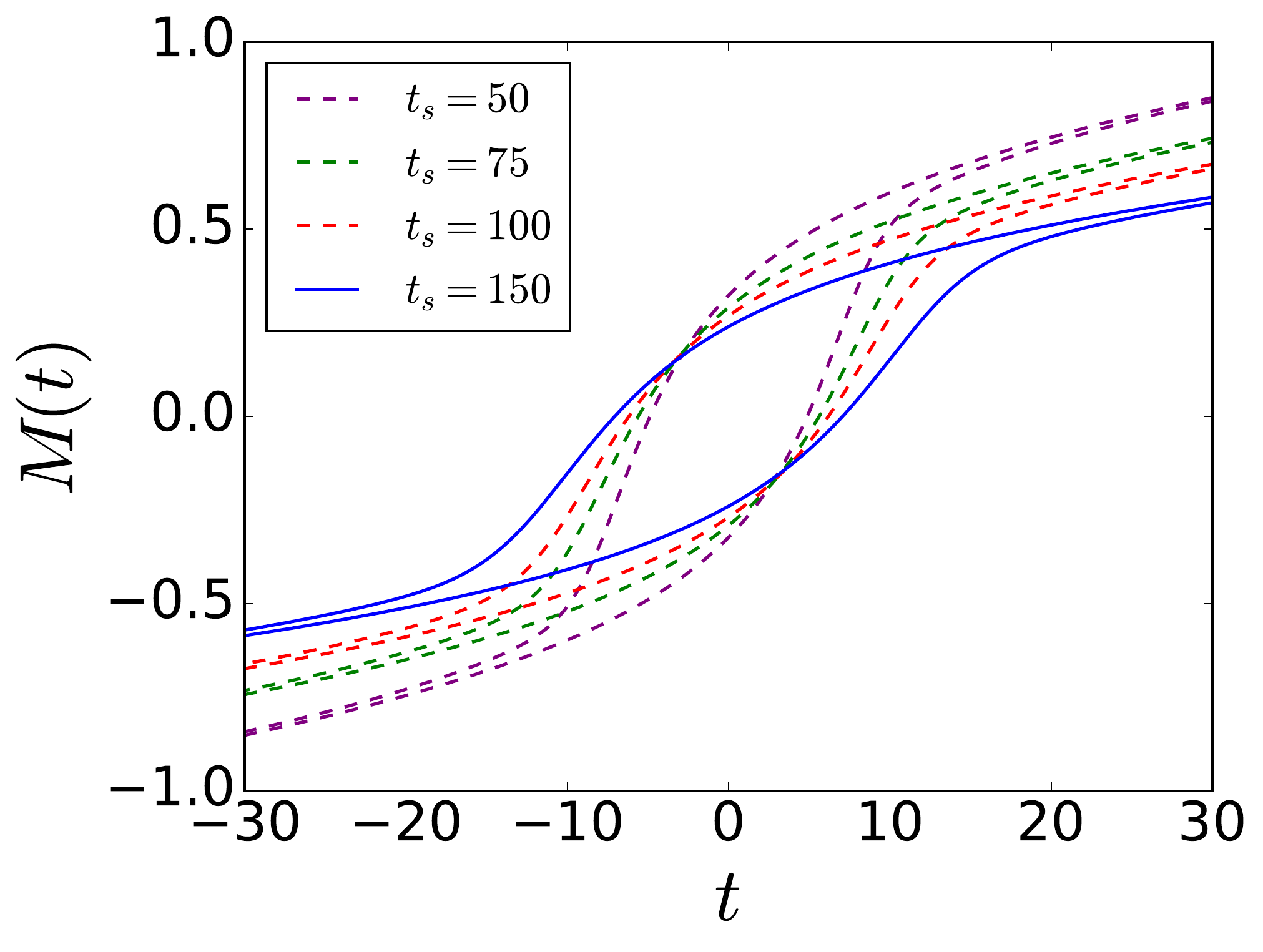} 
\ \includegraphics[width=.475\textwidth]{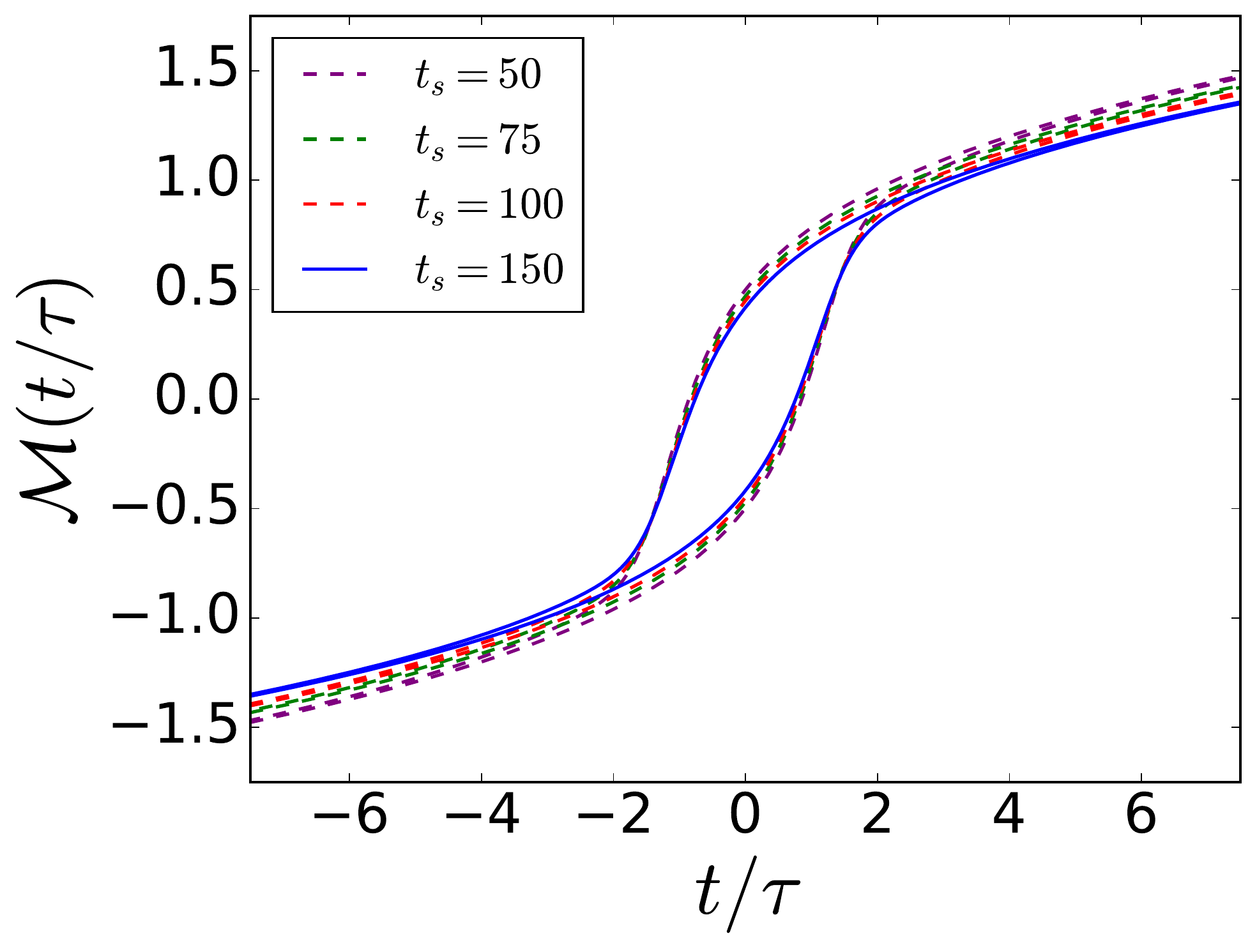}
\caption{\small Numerical analysis of the dynamical magnetisation at $T_c$ during a round-trip protocol in $D=3$ spatial dimensions. \underline{Left panel}: hysteresis loop area for different quench time scales $t_s$. \underline{Right panel}: data collapse and dynamical scaling of the hysteresis area (compare eq~$\eqref{scal-hyst-tc}$) }\label{hyst_rc}
\end{figure}
At the critical temperature $T=T_c$ the hysteresis loop area scales as 
%\textcolor{magenta}{why is there hysteresis for 
%the continuous phase transition?} \blue{that is a great thing of our work man! for a magnetic protocol you can either use a KZ-like theory for the discontinuous phase transition but on the other hand you can define an hysteresis at $\beta_c$ as quantifier of the off-equilibrium! It is not weird since we are in a broken-phase since from the beginning (because of the magnetic field) and moreover it is also confirmed by MC simulations}
%
\begin{equation}\label{scal-hyst-tc}
A\sim \ell^{z-d_{\phi}}\, \mathcal{A} \sim t_s^{\frac{6-D}{6+D}}\, \mathcal{A} \ ,
\end{equation}
where the constant $\mathcal{A}$ reads
% \textcolor{magenta}{Isn't this rather a univeral factor than a scaling fct?}\blue{ Yes you are absolutely right.. That was related to an older version in which ${\cal A}$ was a function of time}
%
\begin{equation}\label{hyst-scal-funct}
\mathcal{A}\! =2\!\int_{-\infty}^{+\infty}\!\!\! \D \bar{t} \int_{-\infty}^{\bar{t}}\!\!\! \D s \, s  
\cosh\!\bigg[\!\int_{s}^{\bar{t}}\!\D u\; \mathfrak{m}^2(u)\bigg]\ .
\end{equation}
Consequently, the dissipated energy (in form of magnetic work) $W$ in $D=3$ spatial dimensions obeys the scaling relation
\begin{equation}\label{work-tc}
W\sim  t_s^{-2/3}  \, \mathcal{A}\ ,
\end{equation}
i.e. the slower the protocol is performed, the less energy is dissipated during the round-trip protocol. This can be
intuitively understood since the system will stay longer in equilibrium for a slow quench.
\begin{figure}[ht]
\centering
\includegraphics[width=.475\textwidth]{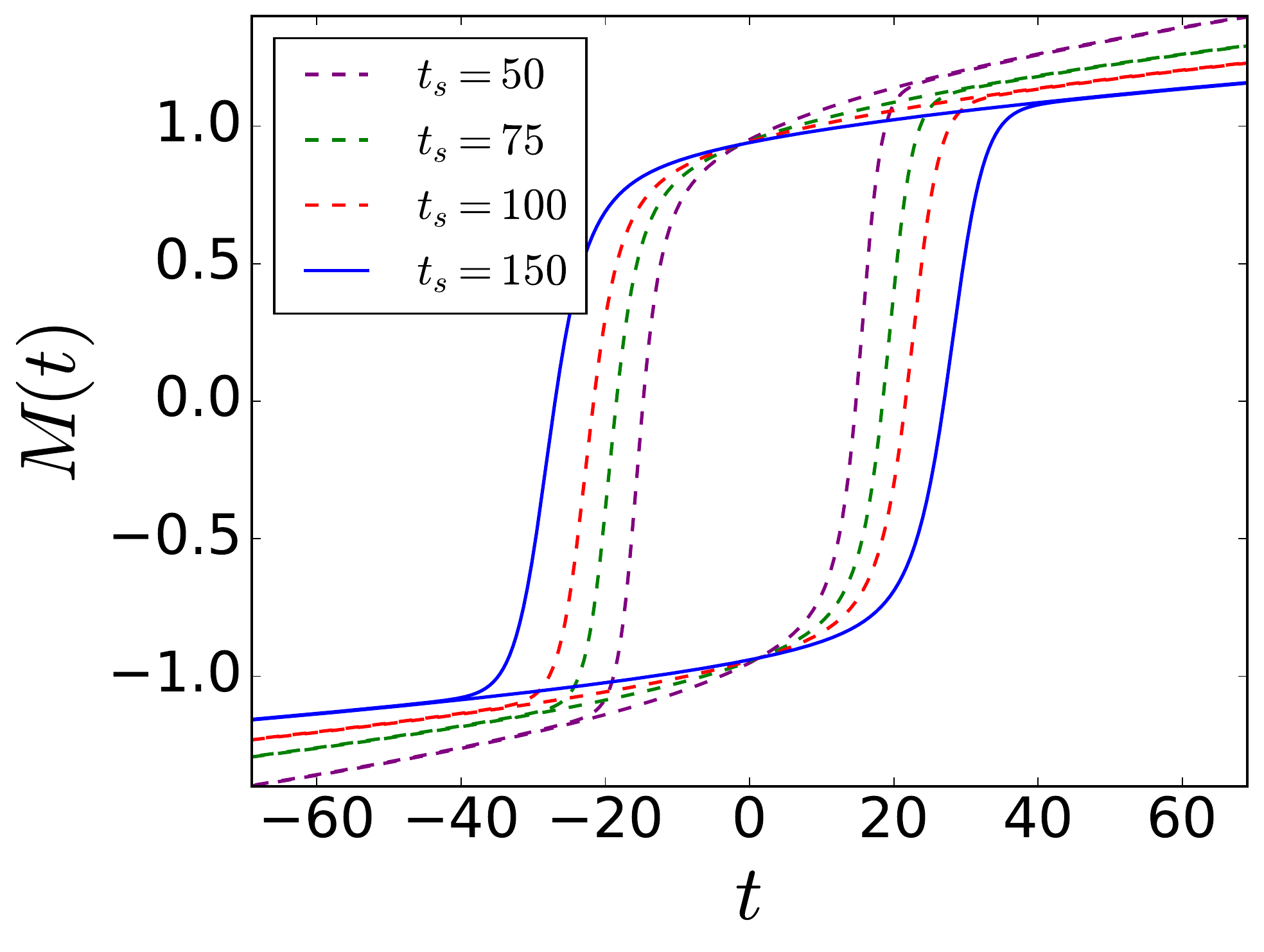} \ 
\includegraphics[width=.475\textwidth]{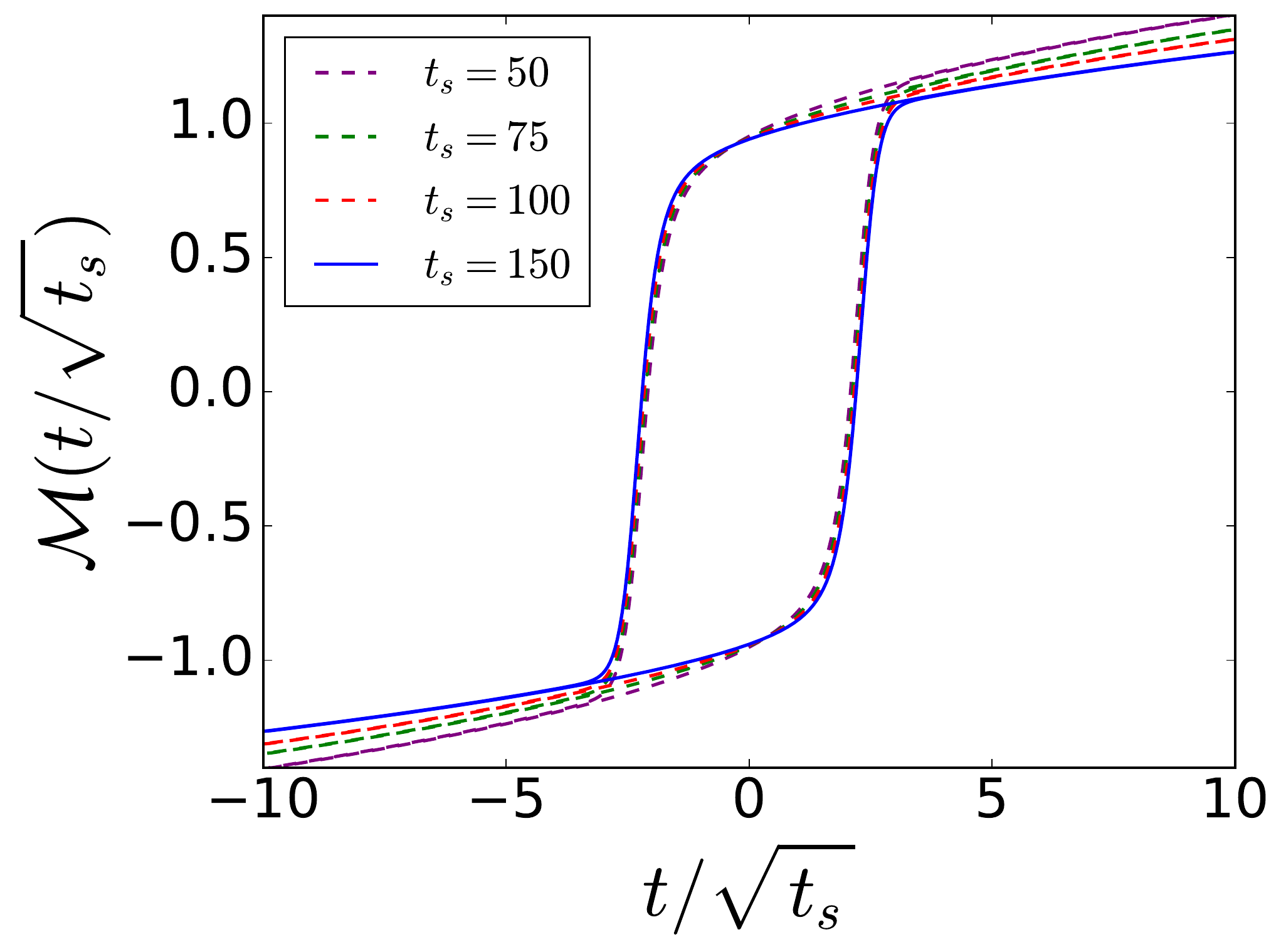}
\caption{\small Numerical analysis of the dynamical magnetisation below $T_c$ ($r=-1$) during a round-trip
protocol in $D=3$ spatial dimensions. \underline{Left panel}: hysteresis loop area for different quench time scales $t_s$. \underline{Right panel}: data collapse and dynamical scaling of the hysteresis area (compare eq~$\eqref{scal-hyst-low}$) }\label{hyst_low}
\end{figure}
Applying the same arguments to the magnetic {\sc fot} below the thermal critical point $T<T_c$ we obtain for 
the hysteresis loop
\begin{equation}\label{scal-hyst-low}
A \sim \ell^{D}_{\text{\sc fot}}\; \mathcal{A}\sim \sqrt{t_s}\, {\cal A}
\end{equation}
where the factor ${\cal A}$ has the same structure as $\eqref{hyst-scal-funct}$ in terms of 
the quantities at low temperature. For the magnetic work, we find the scaling relation
\begin{equation}\label{work-low}
W\sim \ell_{\text{\sc fot}}^{-D} \; \mathcal{A}= \frac{ \mathcal{A}}{\sqrt{t_s}}
\end{equation}
independently from the spatial dimensions $2<D<4$. 
Numerical results for the hysteresis loop area are given in the figures~\ref{hyst_rc} and \ref{hyst_low}, 
respectively for the cases $T=T_c$ and $T<T_c$.

The scaling relations in eqs~(\ref{work-tc},\ref{work-low}) apply beyond the spherical limit $n\To \infty$ and
are in agreement with the numerically obtained scaling behaviour for a 
% 
% correctly predict the scaling behaviour of a 
$3D$ Heisenberg ferromagnet
% , numerically calculated in 
\cite{Vicari-off-16}. Indeed, in the case $T<T_c$, we showed that the dynamics is independent on 
the number of transverse components (see appendix \ref{fot-dyn}) while at the criticality $T=T_c$ the critical
exponents of the $O(n)$ universality classes weakly depend on the {\sc rg} exponent $\eta$ (see table \ref{eta-exp}) so that the large-$n$ limit provides
a reliable guideline. In this sense, we can conclude that {\it the scaling behaviour of the system is 
not modified by considering a finite number of components $n\geq 2$}. The off-equilibrium scaling relations presented 
in this work are briefly summarised in table \ref{scaling-general} for a generic $O(n\geq 2)$ model.
\begin{table}
\centering
\caption{\small Numerical estimations of the critical exponent $\eta$ for different universality classes in three spatial dimensions \cite{Vicari-review}.}
\vspace{.5cm}
\begin{tabular}{|l|c|r|}
\toprule
universality class & $\eta$ & Ref.\\
\midrule
XY & $0.0380(4)$ &\cite{exponents_O(2)}\\
Heisenberg  & $0.0375(5)$ &\cite{exponents_O(3)}\\
$O(4)$ & $0.0365(10)$& \cite{exponents_O(4)}\\
$O(\infty)$ & $0$  & e.g. \cite{Berl52}\\
\bottomrule
\end{tabular}
\label{eta-exp}
\end{table}

%#############################

% \begin{strip}
\begin{table}
\centering
\caption{\small
Off-equilibrium scaling relations for the $O(n)$ universality class with $n\geq 2$ during a magnetic quench $\eqref{eq:quench}$:
at the continuous transition $T=T_c$ with $d_{\phi}=\frac{1}{2}\left(D-2+\eta\right)$, $\nu_h=1/(D-d_{\phi})$ and $z=2-\eta$
and for $T<T_c$.}
\vspace{.25cm}
\begin{tabular}{|l|l|l|}
\toprule
observable &\hspace*{.5cm} scaling $\;T=T_c$ \hspace*{.5cm}&\hspace*{.5cm} scaling $T<T_c$ \hspace*{.5cm}\\
\midrule
magnetisation $M(t)$& $\sim \ell^{-d_{\phi}} \, {\cal M}(t/\tau)$& $\sim t_s^{0} \, {\cal M}(t/\sqrt{t_s})$ \\[.75cm]
trans. susceptibility $ \chi_{\perp}(t)$& $  \sim \ell^{2-\eta} \, \mathcal{X}_{\perp}(t/\tau)$ & $  \sim t_s^{1/2} \, \mathcal{X}_{\perp}(t/\sqrt{t_s})$\\[.75cm]
hysteresis area $A$ & $ \sim \ell^{z-d_{\phi}} \,{\cal A}$ & $\sim t_s^{1/2} \,{\cal A}$ \\
\bottomrule
\end{tabular}
% \vspace{.5cm}
% \begin{flushleft}
% ({\bf b}) \ {\small Scaling relations below the critical temperature $T<T_c$.}
% \end{flushleft}
% % \\[2ex]
% \begin{tabular}{lc}
% \toprule
% observable & scaling $\;T<T_c$ \\
% \midrule
% magnetisation & $M(t)\sim t_s^{0} \, {\cal M}(t/\sqrt{t_s})$ \\[2ex]
% trans. susceptibility & $ \chi_{\perp}(t) \sim t_s^{1/2} \, \mathcal{X}_{\perp}(t/\sqrt{t_s})$ \\[2ex]
% hysteresis area & $A \sim t_s^{1/2} \,{\cal A}$ \\
% \bottomrule
% \end{tabular}
\label{scaling-general}
\end{table}
% \end{strip}
% 
%#############################
%
\section{Summary and conclusion}
We investigated the off-equilibrium scaling arising in classical spin systems due to the presence of a 
time-dependent magnetic field $h(t)=t/t_s$ which 
drives the system  from an initial equilibrium 
state 
%at $t_i<0$ to $t_f>0$ 
across the transition point $h_c=0$ at constant temperature $T\leq T_c$.  In particular, we considered a system with
$O(n)$ symmetry in the large-$n$ limit and in $2<D<4$ spatial dimensions.
We analysed the two distinct scenarios $T=T_c$ and $T<T_c$ which are qualitatively different since the magnetic transition is
continuous at $T=T_c$ and discontinuous for $T<T_c$.

After recalling the general features of the {\sc kz} 
scaling for a continuous transition, we focused on the protocol
below the critical temperature. Here, in the absence of a diverging correlation length, an equilibrium scaling theory is
routinely formulate by referring to the coherence length $\xi_h$ as the characteristic scale. We extended this equilibrium theory 
to the non-equilibrium case by following the general ideas of the {\sc kz} approach. To do so, we
deduced several equilibrium exponents such as e.g. the dynamical exponent $z=D$ for the {\sc fot}, 
needed to formulate the off-equilibrium scaling theory. As a result, thermodynamic 
observables such as the magnetisation or the magnetic susceptibility present dynamical scaling relations
in terms of appropriate off-equilibrium scales. The latter are functions of the quench
time scale $t_s$ and depend on the set of static and dynamic {\sc fot} exponents.
Quite remarkably, these scaling relations have the same structure as
those at $T=T_c$ but with different exponents. 

We then applied this scaling theory to a round-trip protocol, where we proposed the hysteresis area 
as a quantifier of the deviation from 
equilibrium and we derived its scaling behaviour. Moreover, the hysteresis 
can be easily connected to the dissipated magnetic energy by the system during a 
protocol and therefore with an energy cost. 

As mentioned, all results presented in this 
work are derived in the large-$n$ limit. However, we argued that the dynamics of 
the system is not affected by the number of transverse components so that 
our results apply for any finite $n\geq 2$, as confirmed by a comparison with numerical 
studies \cite{Vicari-off-16}.  

Although there are several works on the dynamical off-equilibrium scaling at {\sc fot}s, e.g. in \cite{Vicari-fot-15} where 
thermal quenches are analysed and in \cite{Vic-Dyn1,Vic-Dyn2,Vic-Dyn3} where finite-size scaling in quantum systems
is discussed, the study of non-equilibrium behaviour 
at {\sc fot}s is much less understood and investigated than its continuous counterpart.
We do therefore believe that the simple and clear physical picture that the {\sc kz} mechanism provides, opens new 
perspectives to this field of research which becomes experimentally more and more relevant, especially 
in the light of recent experiments in the area of ultracold atoms, where {\sc fot}s can be generated and studied 
systematically \cite{Land16,Dog16,Nied16,Flo17,Wald18a}.

% We do believe that this work opens a new perspective to the study of non-equilibrium behaviour 
% at first-order transitions, which is much less understood and investigated than its continuous counterpart. 
% We may refer to  \cite{Vicari-fot-15} in which thermal quenches are analysed and to
% \cite{Vic-Dyn1,Vic-Dyn2,Vic-Dyn3} where finite-size scaling in quantum system are discussed. 
% Such descriptions 
% are sought especially in light of recent experiments in ultracold atoms where {\sc fot}s can be generated and studied 
% systematically \cite{Land16,Dog16,Nied16,Flo17,Wald18a}.
%
%
%\textcolor{black}{Moreover, understanding the off-equilibrium scaling properties in analogy with the {\sc kz} scaling theory deepens 
%our knowledge of such transitions for which the classical theory of nucleation \cite{Bray-94} is not sufficient 
%\cite{Zhong3}. This allows us to have a wide view on the scaling at critical points independently from their continuous 
%or discontinuous nature. } \blue{$\star\star$}

A next step might be the extension of the present work to a system with 
inhomogeneities, for which the continuous counterpart is already analysed in the literature e.g. 
\cite{Dragi-Collura, trap-scal1,trap-scal2}. The latter case is closely related to real experimental setups where 
ultracold atomic gases typically do not have a flat density profile due to the effects of a trapping potential
\cite{trap-coldatoms-exp1,trap-coldatoms-exp2,trap-coldatoms-exp3}.\\

\noindent 
{\bf Acknowledgements:} SW is grateful to the LPCT Nancy for their warm hospitality. 
The authors would like to thank Ettore Vicari for his support during the development 
of this work. We appreciate fruitful discussions with Dragi Karevski and Malte Henkel and are thankful
for their critical remarks on the manuscript. \textcolor{black}{Furthermore, we would like to thank the 
referee for his or her careful reading and useful comments on the manuscript.}

% \textcolor{magenta}{\hrule\hrule\hrule}

% \onecolumn

%#################################################################################################
%#################################################################################################

\appendix

\section{Equilibrium limit}\label{equilibrium}

In this appendix we provide further details on how the off-equilibrium scaling behaviour~(\ref{FSS}) and (\ref{scaling-fot})
match their equilibrium counterparts. Therefore, we have to distinguish $T=T_c$ and $T<T_c$.
\subsection{The continuous transition ($T=T_c$)}\label{equilibriumA}
By construction, in equilibrium we can identify the effective mass term of the system with the inverse of the 
square of the instantaneous correlation length obtaining 
\begin{equation}\label{eq-mass}
\mathfrak{m}^2(t/\tau)\propto |t/\tau|^{2\nu_h}\,.
\end{equation}
Notice that this kind of behaviour for $\mathfrak{m}^2(\cdot)$ provides an exponential suppression of the initial 
conditions in $\eqref{eq:M-t}$ and ensures the universality of the scaling behaviour \cite{Sondhi-12}. The 
assumption $\eqref{eq-mass}$ can be easily checked as follows. The equilibrium scaling behaviour can be
defined as the limit $\xi\To \infty$ at fixed $t/\xi^z$ for which the magnetisation $\eqref{eq:M-t}$ shows the behaviour
\begin{equation}
M(t) \sim \xi^{-d_{\phi}}(t) \; M_0 \ ,
\end{equation}
where $M_0$ is a constant. On the other hand the equilibrium matching at $|t|\To \tau$ imposes that
\begin{equation}
M(t) \sim \ell^{-d_{\phi}} \; \mathcal{M}(t/\tau) \ .
\end{equation}
Therefore the scaling function $\eqref{scal-M}$ must satisfy 
\begin{equation}\label{match}
\lim_{t\To -\tau} \mathcal{M}(t/\tau) =M_0\, |t/\tau|^{1/\delta}, 
\end{equation}
with $\delta=d_{h}/d_{\phi}$ being the equilibrium critical exponent. One may verify that by inserting the Ansatz 
$\eqref{eq-mass}$ into $\eqref{scal-M}$, a direct calculation gives the result $\eqref{match}$. 
With the same method we can derive for the transverse susceptibility
\begin{equation}
\lim_{t\To -\tau}\chi_{\perp}(t/\tau) \propto |t/\tau|^{-\gamma}\ ,
\end{equation}
where $\gamma=2\nu_h$ is the equilibrium critical exponent for the system at large-$n$ \cite{LargeN-rev}. 

\subsection{The discontinuous transition ($T<T_c$)}\label{equilibriumB}
Below the critical temperature and close enough to the transition point $h_c=0$, we can approximate 
the equilibrium value of the magnetisation by a constant $M_0=-\sqrt{(r-r_c)/u}$ \cite{Dhar-92}.\footnote{In 
other words, we are assuming that weak magnetic fields do not significantly modify the value of the magnetisation, 
see appendix \ref{spin-wave} for the regime of validity of this approximation.} From this observation eq~$\eqref{dyn-obs}$ gives for the effective mass
\begin{equation}
m^2(t) \approx \frac{h(t)}{M_0}\ ,
\end{equation}
which becomes in the scaling limit
\begin{equation}\label{eq-mass-low}
\mathfrak{m}^2(t/\tau_{\text{ \sc fot}}) = \frac{t/\tau_{\text{ \sc fot}}}{M_0}\,.
\end{equation}
By inserting $\eqref{eq-mass-low}$ in $\eqref{scal-M-low}$ we obtain for the magnetisation 
\begin{equation}
\lim_{t\To -\tau_{\text{\sc fot}}}\mathcal{M}(t/\tau_{\text{ \sc fot}})=M_0 \ .
\end{equation}
Furthermore, a direct calculation of $\eqref{scal-G-low}$ using eq~$\eqref{eq-mass-low}$ leads to 
\begin{equation}
\chi_{\perp}(t)= \frac{M_0}{h(t)} \sim \ell^D_{\text{ \sc fot}}\,\frac{M_0}{t/\tau_{\text{ \sc fot}}}\;,
\end{equation}
in agreement with the general predictions for the transverse susceptibility below the critical temperature \cite{LargeN-rev}.

%#################################################################################################
%#################################################################################################
 
 \section{Dynamics at low temperature $T<T_c$ }\label{fot-dyn}
Here, we analyse the dynamical behaviour of the system in the regime $T<T_c$. The components 
$\phi_{a}(\vec{x},t)$ of the vector field satisfy the equation of motion $\eqref{dynamics}$
 \begin{equation}
 \partial_t \,\phi_{a}(\vec{x},t)= -\left(\nabla^2+m^2(t)\right)\phi_{a}(\vec{x},t) +\delta_{1,a}\, h(t) + \zeta_{a}(\vec{x}, t)\, 
 \end{equation}
from which we notice that all transverse components $a>1$ follow the same evolution. Hence, the {\it number} of transverse 
components does not influence the dynamics and all $n-1$ transverse planes are equivalent. 
It is thus useful to reduce the system to an $O(2)$ model \cite{Dhar-93} where the $2$-component vector field can be 
parametrised as
 \begin{equation}\label{rep-low-temp}
 \phi(\vec{x},t)=|M_0|(1+r(\vec{x},t)) \, \exp[i\theta(\vec{x},t)]\ ,
 \end{equation}
 with a radial fluctuating field $r(\vec{x},t)$ and a dynamical phase $\theta(\vec{x},t) \in [0,2\pi)$. 
 The equation of motion $\eqref{dynamics}$ is decomposed in a set of two coupled equations
 \begin{subequations}
 \begin{align}
 \partial_t\, \theta(\vec{x},t)\, (1+r(\vec{x},t)) = \nabla^2\theta(\vec{x},t) + 2\nabla\theta(\vec{x},t) \nabla r(\vec{x},t) 
 &-\frac{h(t)}{|M_0|} \sin(\theta(\vec{x},t)) +\zeta_{\theta}(\vec{x},t)\ ,\\
\partial_t \, r(\vec{x},t) =\left(1+r(\vec{x},t)\right)\! (\nabla\theta(\vec{x},t))^2 -m^2(t) \left(1+r(\vec{x},t)\right) 
&+\frac{h(t)}{|M_0|}\cos(\theta(\vec{x},t)) +\zeta_r(\vec{x},t)\ ,
\end{align}
\end{subequations}
where now $\vec{x}=(x_1,x_2)$. We have redefined the white Gaussian noise $\eqref{noise}$ as 
\begin{equation}
\zeta_{\theta}(\vec{x},t)\equiv \frac{1}{|M_0|}\big(- \zeta_1(\vec{x},t) \sin[\theta(\vec{x},t)]+\zeta_2(\vec{x},t) \cos[\theta(\vec{x},t)]\big)
\end{equation}
 for the angular motion, and  
 \begin{equation}
 \zeta_{r}(\vec{x},t)\equiv \frac{1}{|M_0|}\big( \zeta_1(\vec{x},t) \cos[\theta(\vec{x},t)]+\zeta_2(\vec{x},t) \sin[\theta(\vec{x},t)]\big)
 \end{equation}
 along the radial direction, both with zero mean and variance $2/M_0^2$.\\
 
  We shall consider the following approximations:
\begin{enumerate}
\item Radial fluctuations are negligible $(1+r)\approx 1$, which provides a good description for weak magnetic fields at 
low temperatures.
\item The angular and radial degrees of freedom are decoupled $\braket{\nabla\theta \cdot \nabla r} =0$.
\item The kinetic term is given by long wavelength modes $|\vec{q}|<q^*$ (see appendix \ref{spin-wave}).
\end{enumerate}

Under these assumptions the evolution of the dynamical phase is described in the Fourier space by 
\begin{equation}\label{phase-evo}
\partial_t \,\theta(\vec{q},t) = -\vec{q}^2\, \theta(\vec{q},t) -\frac{h(t)}{|M_0|}\sin(\theta(\vec{q},t)) +\zeta_{\theta}(\vec{q},t), \qquad |\vec{q}|<q^*
\end{equation}
where $\zeta_{\theta}(\vec{q},t)$ is the Fourier transform of the noise distribution having the same cumulants.

Eq~$\eqref{phase-evo}$ shows two opposite regimes depending on the value of the magnetic field 
\begin{subequations}
\begin{align}\label{CASE1}
\partial_t \,\theta(\vec{0},t) &=  -\frac{h(t)}{|M_0|}\sin(\theta(\vec{0},t)) +\zeta_{\theta}(\vec{0},t)\ , \hspace{2cm}\left|\frac{h(t)}{M_0}\right| \gg (q^*)^2 \simeq 0\ ,
\\[.25cm]
\label{CASE2}
\partial_t \,\theta(\vec{q},t) &= -\vec{q}^2\, \theta(\vec{q},t) +\zeta_{\theta}(\vec{q},t), 
\hspace{3.425cm}\left|\frac{h(t)}{M_0}\right| \ll (q^*)^2 \ .
\end{align}
\end{subequations}
The former case corresponds to the equilibrium limit $|t|\gg \tau_{\text{\sc fot}}$ while the latter describes the off-equilibrium regime $|t|\ll \tau_{\text{\sc fot}}$.
We shall refer to these two regimes with the shorthand notation $h\neq0$ and $h=0$ respectively.
%%%%%%%%%%%%%%%%%%%%%%%%%%%%%%
 \subsection*{Case $h\neq 0$}
We start by analysing the phase dynamics away from the transition point. Here, the time evolution is given by~$\eqref{CASE1}$
\begin{equation}
 \partial_t\, \theta(\vec{0}, t) =  -\frac{h(t)}{|M_0|}\sin(\theta(\vec{0},t)) +\zeta_{\theta}(\vec{0},t) \ . 
 \end{equation}
As we discussed in appendix \ref{equilibrium}, in this regime the value of the magnetisation is not modified significantly. 
It is therefore convenient to consider the Taylor expansion of the phase $\theta(\vec{0},t)$ around the initial value $\theta(\vec{0}, t_i)=\pi$. At the leading order we obtain
\begin{equation}
\partial_t \, \vartheta(t) =  \frac{h(t)}{|M_0|} \, \vartheta(t) +\zeta_{\theta}(t) 
 \end{equation}
 where $\vartheta(t)\equiv \theta(\vec{0},t)-\pi$ and with the formal solution
\begin{equation}
 \vartheta(t)=\int_{t_i}^t \D u\, \exp\bigg[\int_{u}^t \D s\, \frac{h(s)}{|M_0|}\bigg]\, \zeta_{\theta}(u) \ .
 \end{equation}
The mean value of the phase is zero $\braket{\vartheta(t)}=0$ while its variance is
\begin{equation}
\sigma^2(t)\equiv \braket{(\vartheta(t)-\braket{\vartheta})^2} = \frac{2}{M_0^2} \,
\exp\left[\frac{t^2}{|M_0|\, t_s}\right] \; \frac{\sqrt{\pi |M_0| t_s}}{2} \, 
\text{Erfc}\left[\frac{|t|}{\sqrt{|M_0|\, t_s}}\right] \ .
\end{equation}
In the equilibrium limit $|t|\To \tau_{\text{\sc fot}}$ (in 
which the small-angle approximation holds) a straightforward calculation shows
\begin{equation}
 \lim_{t\To-\tau_{\text{\sc fot}}} \sigma^2(t) =  \frac{1}{M^2_0} \, \chi_{\perp}(t) \ .
\end{equation}
The distribution $P(\vartheta, t)$ of the dynamical phase  $\vartheta(t)$ can be 
derived solving the associated Fokker-Plank equation
\begin{equation}
\partial_t \, P(\vartheta,t) = -\frac{\delta}{\delta \vartheta} \left(\frac{h(t)}{|M_0|}\, \vartheta \, P(\vartheta,t)\right) + \frac{\delta^2}{\delta \vartheta^2} \, P(\vartheta,t) \ ,
\end{equation}
with the initial condition $P(\vartheta,t_0)=\delta(\vartheta)$ and having a standard gaussian solution
\begin{equation}
P(\vartheta,t)= \frac{1}{\sqrt{2\pi} \sigma^2(t)} \, \exp\left(-\frac{\vartheta^2}{2\sigma^2(t)}\right)\ .
\end{equation}
In this regime, we conclude that the phase dynamics consists of Gaussian fluctuations ($z=2$) due to the
transverse modes and it leaves the mean value of the magnetisation unchanged.

\subsection*{Case $h=0$ }
Here, we focus on the phase dynamics in the off-equilibrium regime. It is convenient to introduce a finite size 
$L$ in a way that, in absence of anisotropies\footnote{The geometry of the finite-size system influences the scaling 
behaviour, see \cite{Vicari-off-16} for more details. The cubic geometry is the one compatible with the infinite-volume 
considered here.}, the volume of the system is $V=L^D$. In a finite-geometry the dynamical phase and its Fourier modes are related through
\begin{equation}
\theta(\vec{q},t)=\int_V d\vec{x} \; e^{i\vec{q}\,\vec{x}}\, \, \theta(\vec{x},t)\; , \qquad \theta(\vec{x},t)=\frac{1}{V}\sum_q e^{-i\vec{q}\,\vec{x}}\, \theta(\vec{q},t) \; .
 \end{equation}
 The time-evolution of the phase is given by eq~$\eqref{CASE2}$
\begin{equation}
\partial_t\, \theta(\vec{q},t)=-\vec{q}^2\, \theta(\vec{q},t) +\zeta_{\theta}(\vec{q},t) \ ,
\end{equation}
having the formal solution
\begin{equation}\label{theta-h-0}
\theta(\vec{q},t)= e^{-\vec{q}^2\,t}\int_{t_0}^t \D u\, e^{\vec{q}^2 u}\, \zeta_{\theta}(\vec{q},u) \ ,
\end{equation}
where, without loss of generality, we assumed $\theta(\vec{q},t_0)=0$. Using $\eqref{theta-h-0}$ and $\eqref{rep-low-temp}$, we are able to compute the autocorrelation function of the magnetisation
\begin{equation}\label{autocorrM}
\braket{M(t)\; M(s)}\equiv \frac{M^2_0}{V}\int \D\vec{x}\,\D\vec{y} \braket{e^{i\theta(\vec{x},t)-i\theta(\vec{y},s)}}\ .
\end{equation}
At this point,  if we consider the  spatial average of the dynamical phase
\begin{equation}
\Theta(t)\equiv \frac{1}{V}\int_V \D\vec{x}\; \theta(\vec{x},t)
\end{equation}
such that $\theta(\vec{x},t)=\Theta(t)+\mathcal{O}(1/\sqrt{L})$, we may approximate the autocorrelation function $\eqref{autocorrM}$ as \cite{Vicari-off-16}
\begin{equation}
\braket{M(t)\; M(s)}\approx M^2_0\; \braket{e^{i\Theta(t)-i\Theta(s)}} =M_0^2 \, \exp\left[-\frac{|t-s|}{M^2_0\, V}\right]
\end{equation}
 from which we deduce that the autocorrelation time is of the order of $L^D$ implying that the dynamical exponent is $z=D$.

%#################################################################################################
%#################################################################################################

\section{Spin-wave approximation}\label{spin-wave}
We consider the expression $\eqref{eq:GT-t}$ for the transverse correlation function
\begin{equation}\label{corr}
G_{\perp}(\vec{q},t)=2\int_{t_0}^t \D u \, \exp\bigg[-2\int_{u}^t \D s\, (\vec{q}^2+m^2(s))\bigg].
\end{equation}
As argued in appendix \ref{equilibrium}, below the critical temperature and for weak magnetic fields $h\simeq 0$ 
the magnetisation can be approximated by its equilibrium value $M_0$. In this approximation scheme, we obtain the 
estimate of the mass term $\eqref{eq-mass-low}$. We already know that this approximation breaks down when
$|t|\leq \tau_{\text{\sc fot}}$, where the magnetisation cannot be considered as constant anymore. However, the naive use of the Ansatz 
$\eqref{eq-mass-low}$ permits to compute the value of the transverse correlation function explicitly
\begin{equation}
 G_{\perp}(\vec{q},t) \approx  e^{s^2} \, \sqrt{\frac{\pi}{2}}\, \sqrt{|M_0|\, t_s}\,
 \Big( \text{Erfc}(s(\vec{q},t))-\text{Erfc}(s(\vec{q},t_0))\Big) \ ,
  \end{equation}
where $s(\vec{q},t)\equiv (\vec{q}^2\, |M_0|\, t_s + |t|)/\sqrt{|M_0|\,t_s}$. For large $s\gg 1$ the equilibrium propagator is recovered
\begin{equation}
\lim_{s\To\infty} G_{\perp}(s(\vec{q},t)) = \frac{1}{\vec{q}^2 + h(t)/M_0} = \frac{1}{\vec{q}^2 +m^2(t)}\ .
\end{equation}
Notice that the limit $s\to\infty$ does not necessarily imply $|t/\tau_{\text{\sc fot}}|\To \infty$, i.e. the equilibrium
limit. Indeed, considering large momenta, the system appears at any time in equilibrium. We shall therefore consider for 
the non-equilibrium dynamics only long-wavelength fluctuations $|\vec{q}|<q^*$ while we assume that modes 
$|\vec{q}|>q^*$ are always in equilibrium. 
The boundary value $q^*$ which separates the two regimes can be obtained imposing the condition $s(q^*, t=0) \sim 
\mathcal{O}(1)$ \cite{Dhar-92}. From the latter we have the estimation
\begin{equation}
q^* \propto t_s^{-1/4} \ .
\end{equation}

\section{Numerical Implementation}\label{numerics}
The dynamical eqs (\ref{dyn-obs},\ref{dyn-obsG},\ref{time-eq-state}) can be numerically solved using an iterative 
method, see e.g. \cite{Paes03}.
We first divide the time-window of the protocol $t_s=t_f-t_i$ into ${\cal N}$ parts
\begin{equation}
k= \frac{t_s}{\cal N}\,, \qquad k\ll t_s
\end{equation}
and we consider the discretised time variable $t^{(j)}= t_i + k j $, $j=0,\dots, {\cal N}$. Any function of time $f(t)$
may then be written in discretised version as a ${\cal N} + 1$-vector
\begin{equation}
f(t) \; \longmapsto\; f= \begin{pmatrix} f_0 , \dots , f_{\cal N} \end{pmatrix}\,, \qquad f_j= f\left(t^{(j)}\right)
\end{equation}
and the time derivative may be replaced by a finite difference
\begin{equation}
\frac{d}{dt}\, f(t) \; \longmapsto \; \frac{f_{j+1}-f_{j}}{k}\,, \quad \forall j\,.
\end{equation}
We then have to consider the integral over the momenta of the transverse correlation function
\begin{equation}
\int_{\vec{q}} G_{\perp}(\vec{q}, t) =\frac{\Omega_D}{(2\pi)^D} \int_0^{\Lambda} \D q \, q^{D-1}\, G_{\perp}(q, t)
\end{equation}
since the transverse correlation function depends only on $|\vec{q}|$. To estimate this integral, we discretise the momenta
\begin{equation}
\kappa= \frac{ \Lambda}{{\cal N}_q}\ , \qquad \kappa \ll \Lambda
\end{equation}
so that the discretised momentum is $q= \kappa z$, $z=0,\dots,{\cal N}_q$. Momentum integrations can then be
evaluated as
\begin{equation}
\frac{\Omega_D}{(2\pi)^D} \int_0^{\Lambda} \D q \, q^{D-1}\, G_{\perp}(q, t) \approx \frac{\Omega_D}{(2\pi)^D}\, \kappa \left( \sum_{z=1}^{{\cal N}_q-1} \left(\kappa z\right)^{D-1}\, G_{\perp}(\kappa \, z, t) + \frac{1}{2} \Lambda^{D-1} \, G_{\perp}(\Lambda, t)\right)
\end{equation}
 using the extended trapezoidal rule \cite{trapez}. The discretised versions of the eqs 
 (\ref{dyn-obs},\ref{dyn-obsG},\ref{time-eq-state}) then read
 \begin{subequations}
 \begin{align}
 M_{j+1}&= M_j + k \left[h_j -s_j \, M_j \right]\\[.4cm]
g_{z,j+1}&=  g_{z,j}-2 k\left[\left( (\kappa z)^2 + s_j \right)\, g_{z,j} -1\right]
\label{eq3}\\
s_j&= r + u \bigg[ M_j^2 +  \frac{\Omega_D}{(2\pi)^D}\, \kappa \bigg( \sum_{z=1}^{{\cal N}_q-1} \left(\kappa z\right)^{D-1}\, g_{z,j} + \frac{\Lambda^{D-1}}{2}  \, g_{{\cal N}_q, j} \bigg) \bigg]
\end{align}
\end{subequations}
with $M_j\equiv M(t^{(j)})$, $g_{z,j}\equiv G_{\perp}(\kappa z, t^{(j)})$, $s_j\equiv m^2(t^{(j)})$ and $h_j= t^{(j)}/t_s$. This set of algebraic equation can be solved iteratively starting from the initial equilibrium constraints
\begin{equation}
s_0= \frac{h_0}{M_0}\,, \qquad g_{z,0}= \frac{1}{\kappa^2 z^2 + s_0}
\end{equation}
with $M_0$ given by $\eqref{eq3}$ for $j=0$. For simplicity, we focus on $D=3$ and we set $u=1$. The thermal
critical coupling then reads $r_c=-1/2\pi^2\simeq -0.051$ and we can explore the cases $r=r_c$ and $r<r_c$ respectively.

{\small 
\bibliographystyle{condmat.bst}
\bibliography{mybib}
}

\end{document}